\documentclass[journal,twoside]{IEEEtran}
\usepackage{cite}
\usepackage[cmex10]{amsmath}
\usepackage{amssymb,amsfonts}
\usepackage{amsthm}
\usepackage{float}
\usepackage{graphicx}
\usepackage{textcomp}
\usepackage{mathrsfs}
\usepackage{amsmath}
\allowdisplaybreaks[4]
\usepackage{bbm}
\usepackage{xcolor}
\usepackage{marvosym}
\usepackage[tight,footnotesize]{subfigure}
\newcommand{\mr}{\mathrm}
\usepackage{url}
\newcommand{\bb}{{\rm BB}}
\newcommand{\rf}{{\rm RF}}
\newcommand{\tabincell}[2]{\begin{tabular}{@{}#1@{}}#2\end{tabular}}
\usepackage{tablefootnote}
\usepackage{threeparttable}
\normalsize
\usepackage{caption}
\floatstyle{ruled}
\newfloat{algorithm}{tbp}{loa}
\providecommand{\algorithmname}{Algorithm}
\floatname{algorithm}{\protect\algorithmname}

\ifCLASSINFOpdf
\else
\fi
\begin{document}
%\history{Date of publication xxxx 00, 0000, date of current version xxxx 00, 0000.}
%\doi{10.1109/ACCESS.2023.1120000}

\title{Performance Analysis of In-Band-Full-Duplex Multi-Cell Wideband IAB Networks}
\author{Junkai Zhang, \IEEEmembership{Member, IEEE,} 
and Tharmalingam Ratnarajah, \IEEEmembership{Senior Member, IEEE}

\thanks{J. Zhang is with School of Mathematics and Statistics, Xi'an Jiaotong University, Xi'an, 710049, China
(e-mail: jk.zhang@xjtu.edu.cn).}
\thanks{T. Ratnarajah is with the Institute for Imaging, Data and Communications (IDCOM), The University of Edinburgh, Edinburgh, EH9 3BF, UK
(e-mail: t.ratnarajah@ed.ac.uk).}
\thanks{Junkai Zhang was with the Institute for Imaging, Data and Communications, The University of Edinburgh and he is now working with the School of Mathematics and Statistics, Xi'an Jiaotong University.
}
\thanks{This work was supported by the U.K. Engineering and Physical Sciences Research Council (EPSRC) under Grant EP/T021063/1.}
\thanks{This work is based on the Ph.D. dissertation \cite{mythesis} of Junkai Zhang.}
\thanks{Corresponding author: Junkai Zhang (e-mail: jk.zhang@xjtu.edu.cn).}}

\markboth
{Zhang \MakeLowercase{\textit{et al.}}: Performance Analysis of In-Band-Full-Duplex Multi-Cell Wideband IAB Networks}
{Zhang \MakeLowercase{\textit{et al.}}: Performance Analysis of In-Band-Full-Duplex Multi-Cell Wideband IAB Networks}

\maketitle

\begin{abstract}
This study analyzes the performance of 3rd Generation Partnership Project (3GPP)-inspired multi-cell wideband single-hop backhaul millimeter-wave-in-band-full-duplex (IBFD)-integrated access and backhaul (IAB) networks using stochastic geometry. We modeled the wired-connected Next Generation NodeBs (gNBs) as the Matérn hard-core point process (MHCPP) to meet real-world deployment requirements and reduce the cost caused by wired connections in the network. We first derive association probabilities that reflect the likelihood that typical user-equipment is served by a gNB or an IAB-node based on the maximum long-term averaged biased-received-desired-signal power criteria. Furthermore, by leveraging the composite Gamma-Lognormal distribution, we derived the closed-form signal to interference plus noise ratio coverage, capacity with outage, and ergodic capacity of the network. To avoid underestimating the noise, we consider the sidelobe gain on the inter-cell interference links and analog-to-digital converter quantization noise. Compared with half-duplex transmission, the numerical results show an enhanced capacity with outage and ergodic capacity provided by the IBFD under successful self-interference cancellation. We also study how the power bias and density ratio of the IAB-node to gNB and the hard-core distance can affect system performance.
\end{abstract}


%\titlepgskip=-21pt

\maketitle

\section{Introduction}
\label{sec:introduction}
\IEEEPARstart{M}{illimeter}-wave (mmWave) communications have been provided to satisfy the urgent demand for large available bandwidth in beyond fifth-generation (B5G) communications, as the number of wireless devices increases sharply in current networks \cite{MAG,tong}. Although a large available bandwidth of 30-300 GHz can significantly increase the capacity of the wireless network, the path loss becomes more severe than that in microwave communications because the effective antenna aperture is proportional to the wavelength. Therefore, dense base station (BS) deployment with large-scale antenna arrays is required. For hardware and power efficiency, fully-connected/subarray hybrid beamforming is used for mmWave communication \cite{7448873,7434599}.

To further explore the benefits of the mmWave spectrum, in-band-full-duplex (IBFD) transmission, where transmission and reception occur at the same time and frequency band, has been proposed \cite{8246856,7562572}. Compared with half-duplex (HD) transmission, the IBFD scheme provides twice the spectral efficiency (SE) and half the latency \cite{luo}. However, dealing with self-interference (SI), which can be more than 100 dB higher than the desired signal, is the biggest challenge in realizing IBFD transmission. Fortunately, a combination of antenna \cite{9431171}, analog \cite{luo,9500615}, and digital \cite{Zhan2012} SI cancellation (SIC) can achieve satisfactory performance in mmWave IBFD networks.

Moreover, integrated access and backhaul (IAB) networks have been studied in the technical specifications--TR 38.874 (Rel. 16) provided by the 3rd Generation Partnership Project (3GPP) \cite{3gpp}, where wireless backhaul is introduced instead of fiber connections. Plug-and-play deployment \cite{3gpp}, multi-hop networks \cite{8252876}, and performance analysis \cite{8493520} of IAB networks have been used in recent research. Although the authors in \cite{9040265} concluded that all-wired networks (i.e., all BSs connected to the core network by fiber directly) provide better performance than IAB networks, IAB networks provide a good solution for reducing the deployment cost in mmWave communications. Although we designed single-cell wideband single-hop backhaul mmWave-IBFD-IAB networks in \cite{MAG,Zhan2012}, more profound studies on the performance analysis of multi-cell wideband single-hop backhaul mmWave-IBFD-IAB networks are lacking in the literature, which is the main subject of this work.

In previous studies \cite{8882288,tong,8493520,7110547}, stochastic geometry was used to analyze the performance of IAB networks. However, none of these studies considered IBFD transmission and hybrid beamforming, except for \cite{tong}. In addition, in terms of the distribution of small fading, only \cite{8493520} considered the Nakagami-$M$ distribution, which is more suitable for mmWave communications than the Rayleigh distribution because the Rayleigh fading model for sub-6 GHz communications is predicated on a large amount of local scattering, which is not the case for mmWave communications \cite{6515173}. Moreover, the shadowing effect was lacking in most studies. Only \cite{7110547} considers both small fading and path loss with the shadowing effect; however, the Rayleigh distribution is used to model the small fading rather than the Nakagami-$M$ distribution. Therefore, to analyze a more general mmWave-IAB network suitable for beyond 5G communications, we consider IBFD, hybrid beamforming, Nakagami-$M$ small fading, and lognormal shadowing effects in this work.

{Stochastic geometry is a powerful tool for analyzing the performance of wireless networks such as mmWave cellular networks \cite{6932503}, heterogeneous cellular networks \cite{6287527}, and unmanned aerial vehicle (UAV) networks \cite{8335329}. However, to the best of our knowledge, work on IAB systems with stochastic geometry is insufficient. For instance, the performance of mmWave-HD-IAB systems was analyzed in \cite{8882288} with two resource allocation strategies at Next Generation NodeBs (gNBs), i.e., integrated resource allocation and orthogonal resource allocation. The authors in \cite{tong} studied the performance of mmWave-IBFD-IAB systems in wireless edge caching. Bandwidth partitioning for mmWave-HD-IAB systems with performance analysis by stochastic geometry was studied in \cite{8493520}. A multi-cell model of a single-tier mmWave-HD-IAB system was proposed in \cite{7110547}. However, in the aforementioned stochastic geometry-related IAB networks studies, only the work in \cite{7110547} considered the shadowing effect in mmWave communications. Moreover, Nakagami-$M$ small fading was only introduced in \cite{8493520}. }

Meanwhile, in the real world, BSs are not entirely deployed randomly or regularly, which results in the deterministic or Poisson point process (PPP) model becoming insufficient. Therefore, non-Poisson models are preferred for modeling real-world BS deployment. For analytical tractability, the authors of \cite{8648502} suggested a method that uses PPP to approximate the signal-to-interference-ratio meta-distribution for non-Poisson networks, which can provide good accuracy. In this work, to model the deployment of wired-connected gNBs, that is, IAB donors, the Matérn hard-core point process (MHCPP) was applied, where a hard-core distance (i.e., the minimum distance between two gNBs separated in the network) was utilized. The MHCPP model reflects the repulsion between gNBs because gNBs act as wired-anchored BSs. The two gNBs should not be too close or overlapping. Moreover, while providing satisfactory service quality, MHCPP generates a lower density of gNBs than PPP, which avoids the high cost of wired connections. Although MHCPP has been applied to the modeling of UAVs \cite{9350211}, multi-cell IBFD cellular networks \cite{9258892}, and cloud radio access networks \cite{7415954}, to the best of our knowledge, this is the first study to introduce MHCPP into the analysis of multi-cell mmWave-IBFD-IAB networks.

In this study, we evaluate the performance of multi-cell wideband single-hop backhaul mmWave-IBFD-IAB networks using stochastic geometry. The main contributions of this study are summarized as follows. {A comparison between the existing stochastic geometry-based studies and our work is summarized in Table.~\ref{sgc}.}
\begin{table*}[!t]
\renewcommand{\arraystretch}{0.8}
%\extrarowheight %as needed to properly center the text within the cells
\caption{Comparison between existing stochastic geometry-based studies and our work}
\label{sgc}
\centering
\begin{threeparttable}
 {\begin{tabular}{|c|c|c|c|c|c|c|c|c|c|}
\hline
& mmWave & IBFD&IAB&Hybrid Beamforming&Nakagami-$M$&Lognormal&MHCPP&Sidelobe Gain&Quantization Noise\\
\hline
\cite{tong}&$\surd$&$\surd$&$\surd$&$\surd$&&&&$\surd$&\\
\hline
\cite{7434599}&$\surd$&&&$\surd$&&$\surd$&&$\surd$&\\
\hline
\cite{8493520}&$\surd$&&$\surd$&&$\surd$&&&&\\
\hline
\cite{8882288}&$\surd$&&$\surd$&&&&&&\\
\hline
\cite{7110547}&$\surd$&&$\surd$&&&$\surd$&&&\\
\hline
\cite{6932503}&$\surd$&&&&$\surd$&&&&\\
\hline
\cite{8335329}&$\surd$&&&&$\surd$&&$\surd$&&\\
\hline
\cite{9258892}&&$\surd$&&&&&$\surd$&&\\
\hline
\cite{7817893}&&$\surd$&$\surd$&&&&&&\\
\hline
\cite{7448962}&$\surd$&&&&$\surd$&&&$\surd$&\\
\hline
\cite{7434598}&$\surd$&&&$\surd$&&$\surd$&&$\surd$&\\
\hline
\cite{7857035}&&&&&$\surd$&$\surd$&&&\\
\hline
Our work&$\surd$&$\surd$&$\surd$&$\surd$&$\surd$&$\surd$&$\surd$&$\surd$&$\surd$\\
\hline
\end{tabular}}
\end{threeparttable}
\end{table*}

\begin{itemize}
    \item Unlike previous studies on mmWave communications using stochastic geometry, which do not consider hybrid beamforming, we analyzed the performance with energy-efficient subarray-based hybrid beamforming because mmWave communications need to adopt large-scale antenna arrays. Hybrid beamforming has lower hardware cost and power consumption than transitional full digital beamforming { because the number of RF chains required is reduced}.
    \item Unlike prior studies on IAB networks, which use PPP to model the location of gNBs, we use MHCPP to meet the real-world deployment scenario in which BSs are deployed in the non-Poisson model. In this manner, wired-connected gNBs, which act as anchored BSs, can be deployed without overlapping or being too close. In addition, MHCPP produces fewer gNBs than PPP, which reduces the cost of deploying wired gNBs. We use the probability generating functional (PGFL) of PPP to estimate the mean interference from gNBs in the backhaul link and gNB-associated access link, which makes the analysis more tractable.
    \item Instead of analyzing the performance of mmWave-HD-all-wired networks, we consider mmWave-IBFD-IAB networks, which satisfy the requirement of enhanced communication performance and dense deployment with less cost in beyond 5G communications. The IAB-nodes operate in the IBFD mode for simultaneous transmission and reception. Assuming successful SIC, only the residual SI (RSI), modeled as Gaussian noise, is considered, whose power is modeled as a fraction of the transmit power {\cite{tong,7817893,Zhan2012}}.
    \item To accurately model the channel and path loss for mmWave communications, we consider Nakagami-$M$ small fading and include the lognormal shadowing effect in the path loss model, which is different from most of the existing literature on mmWave communications using stochastic geometry. For tractability, we introduced the closed-form statistical property of the composite Gamma-Lognormal (GL) distribution to derive the analytical results.
    \item To avoid underestimation of noise and interference, we consider the sidelobe gain for the inter-cell interference links and analog-to-digital converter (ADC) quantization noise.
\end{itemize}

Numerical results show that by tuning the bias ratio of the IAB-node to gNB, signal-to-interference-plus-noise-ratio ($\mr{SINR}$) coverage yields a convex-like curve at a fixed $\mr{SINR}$ threshold. With a selected bias ratio (e.g., 0 dB), the capacity with outage of the IBFD scheme outperforms that of the HD scheme, regardless of the RSI power. In addition, for our default simulation setting, at $\mr{SINR}=0$ dB, the ADC quantization noise effect becomes negligible at a resolution of approximately five bits, which emphasizes the feasibility of using a low resolution ADC in our proposed network. Moreover, under the same gNB density, the network ergodic capacity obtained by deploying gNBs with the PPP-based model is less than that obtained by deploying gNBs with the MHCPP-based model; the larger the hard-core distance, the higher the ergodic capacity. In addition, with successful SIC, the ergodic capacity increases as the density ratio of the IAB-node to the gNB increases.

\textit{Notations:} $\mathcal{B}, \mathbf{B}, \mathbf{b}$, $b$ represent the set, the matrix, the vector, and the scalar, respectively. $\mathbf{B}^{H}, \mathbf{B}^{-1}$, and $\mathbf{B}^T$ are the Hermitian, inverse, and transpose of $\mathbf{B}$, respectively. $\mr{blkdiag}\left[\mathbf{B}_1,\mathbf{B}_2\right]$ is the block diagonal matrix formed by matrices $\mathbf{B}_1$ and $\mathbf{B}_2$. $\left\|\cdot\right\|_F$ denotes the Frobenius norm. $|b|$ is the norm of $b$. $||\mathbf{b}||$ is the norm of $\mathbf{b}$. $C_n^k$ is the possible combinations of $k$ elements out of a group of $n$ elements where order does not matter. $\mathcal{N}(m,n)$ represents a normal distribution with mean $m$ and variance $n$. $\mr{ln}\mathcal{N}(m,n)$ denotes a lognormal distribution with parameters $m$ and $n$. $\mathcal{U}[a,b]$ represents a uniform distribution between $a$ and $b$.

\section{System Model}
\subsection{IAB Network Topology}
We consider the downlink of a  standalone single-hop backhaul IAB network with a spanning tree (ST) topology consisting of the following nodes:
\begin{itemize}
    \item IAB donor: This is also called a gNB, which is a single logical node. The gNB connects to the next-generation core network using fibers and wirelessly transmits signals to other types of nodes in the network.
    \item IAB-node: This node operates in the IBFD manner, contributes SI from its transmitter to its receiver, and uses wireless links to communicate with others. Owing to the single-hop backhaul and ST topology, an IAB-node has only one parent gNB for backhauling.
    \item User-equipment (UE): It receives the desired signal from its serving gNB or IAB-node.
\end{itemize}
The backhaul link is the wireless link between the gNB and the IAB-node, whereas the access link is between the UE and the gNB or the IAB-node. More details of the 3GPP architecture can be found in our recent study \cite{MAG}. An illustration of the proposed model is shown in Fig.~\ref{SA}. In this study, the gNBs and IAB-nodes formed two different tiers. Multi-hop backhaul IAB networks will not be considered because of the challenge of network configuration, and the feasibility of using stochastic geometry in this scenario \cite{8882288}.

\subsection{Spatial Arrangement}
\begin{figure}[t!]
\centering
\subfigure[]{
\includegraphics[width=\columnwidth]{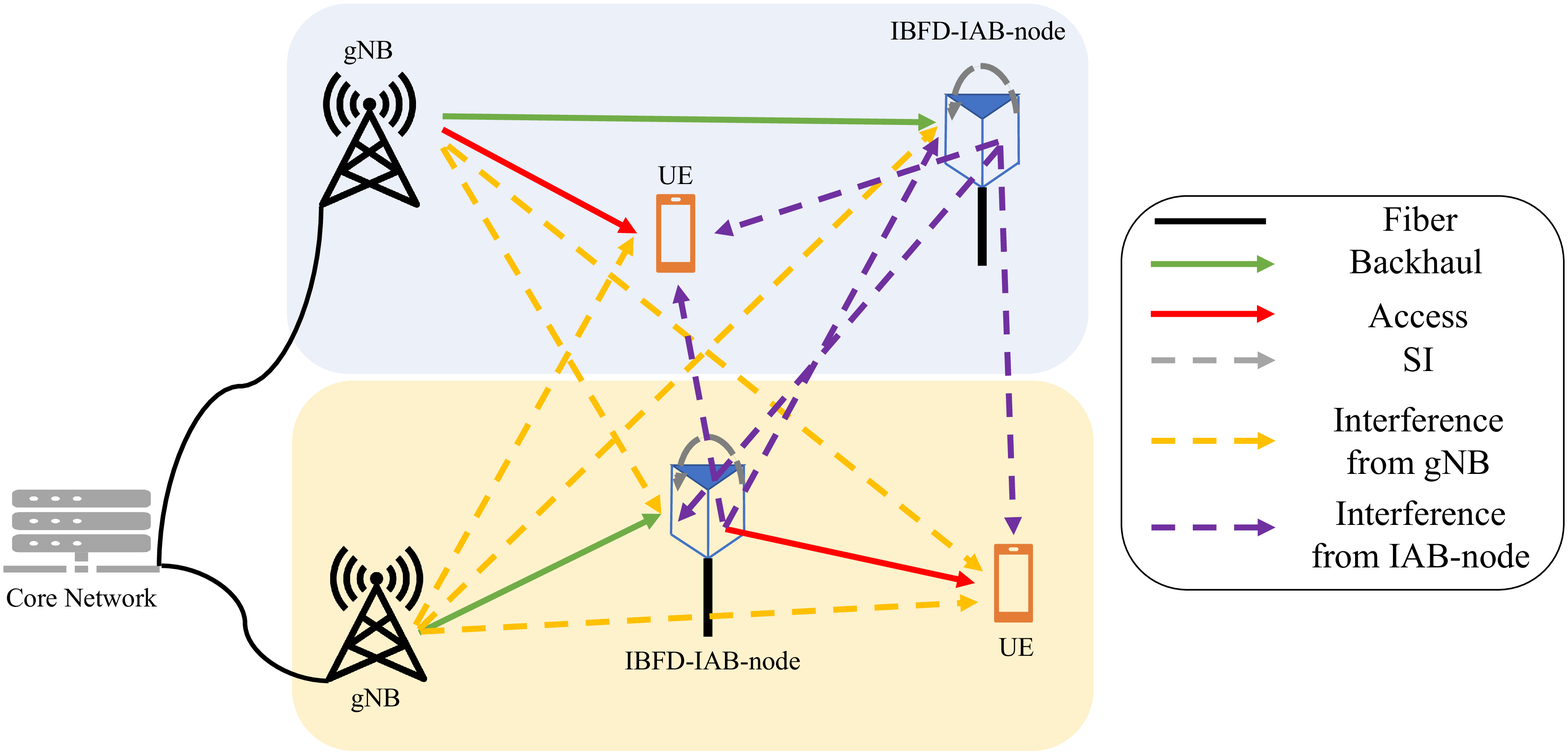}\label{SA}}
\subfigure[]{
\includegraphics[width=\columnwidth]{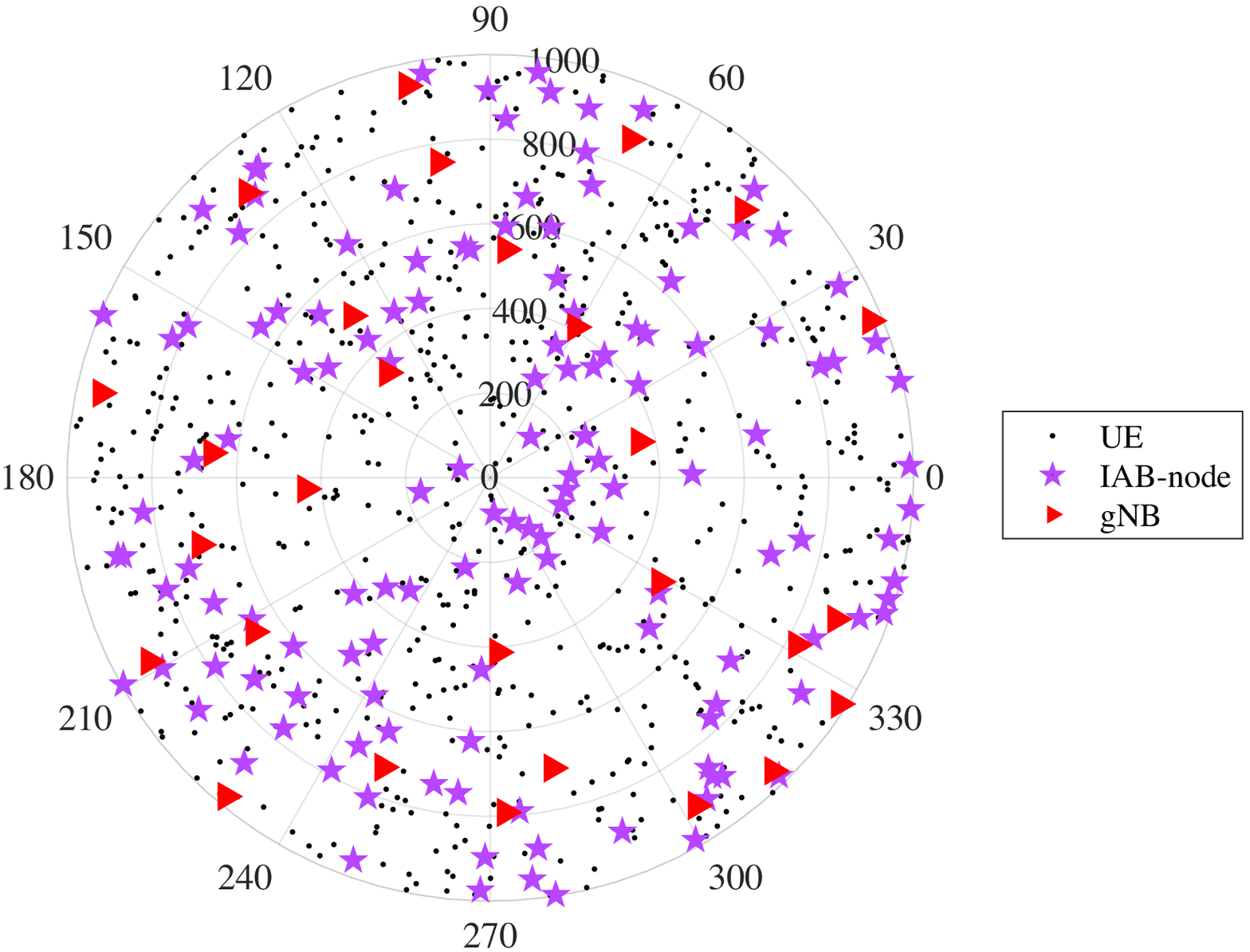}\label{real}}
\caption{(a) Illustration of a multi-cell single-hop backhaul mmWave-IBFD-IAB network. (b) A realized mmWave-IBFD-IAB network. $R_0=1000$m, $\lambda_\mr{m}=1\times10^{-5}/\text{m}^2$, $\lambda_\mr{s}=4\times10^{-5}/\text{m}^2$, $\lambda_\mr{u}=2\times10^{-4}/\text{m}^2$, $\xi=100$m.}
\end{figure}

There are two types of {MHCPP}: both are generated by thinning the parent {PPP}. In Type I, a point and its neighbors whose distance is less than the hard-core distance are removed from the parent {PPP}. In type II, each point is assigned a random number between 0 and 1 as their marks; only points that have larger marks and are within the hard-core region of their neighbors are removed \cite{haenggi_2012}. In this study, we utilize {MHCPP} Type II\footnote{Under the same parent {PPP} density, the thinning rule of {MHCPP} type II is more relaxed than that of {MHCPP} Type I, which can provide a higher density of gNB to satisfy the dense deployment condition for {mmWave} communications.} to capture the location of gNBs. Consider that gNBs are distributed according to an {MHCPP} $\Phi_\mathrm{m}$ with density $\lambda_\mathrm{m}$ and hard-core distance $\xi$ on $\mathbb{R}^2$, where the density is defined by the dependent thinning of a parent homogeneous {PPP} $\widetilde{\Phi}_\mathrm{m}$ with density of $\widetilde{\lambda}_\mathrm{m}$. Given that the probability of an arbitrary point in the parent {PPP} is retained for {MHCPP} Type II as
\begin{equation}
    \rho=\frac{1-e^{-\pi \xi^2\widetilde{\lambda}_\mathrm{m}}}{\pi \xi^2\widetilde{\lambda}_\mathrm{m}},
\end{equation}
where $\lambda_\mathrm{m}=\rho\widetilde{\lambda}_\mathrm{m}$  \cite{5934671}. The {IAB}-nodes and {UE} are drawn from two independent homogeneous {PPP} $\Phi_\mathrm{s}$ and $\Phi_\mathrm{u}$ with densities $\lambda_\mathrm{s}$ and $\lambda_\mathrm{u}$ on $\mathbb{R}^2$, respectively. Our analysis was performed on a circular region with a radius $R_0$. A realization of such a network is illustrated in Fig.~\ref{real}.

\subsection{Propagation Model}
\subsubsection{Blockage Model}
At {mmWave} frequencies, the signals are easily blocked by obstacles in the propagation environment, which results in the coexistence of line-of-sight (LoS) and non-line-of-sight (NLoS) propagations. Blockages typically form a process of random shapes, for example, a Boolean scheme of rectangles \cite{6840343}. In this section, the probability of being {LoS} $p_\mathrm{L}(r)$ or {NLoS} $p_\mathrm{N}(r)$ transmission is modeled by a two-state exponential blockage model with respect to the blockage density constant $\epsilon$ and the link distance $r$ \cite{7448962}, which is expressed as
\begin{equation}
\left\{
\begin{aligned}
    \quad& p_\mr{L}(r)=e^{-\epsilon r},\\ &p_\mr{N}(r)=1-e^{-\epsilon r}.
    \end{aligned}
    \right.
\end{equation}

Note that we adopted the above model for {IAB}-nodes because gNBs can be located well above ground level because of the role of anchored BSs and are usually placed on open ground with fewer obstacles to increase the likelihood of {LoS} links. For tractability, we assume {LoS} transmission only for the gNB to ease the lack of transformation properties of {MHCPP}, such as the {PGFL}. Moreover, we consider the independent blockage effects between the links in this study.

Given the blockage model, we can treat the {PPP} {IAB}-nodes $\Phi_\mathrm{s}$ as the superposition of {PPP} {LoS} {IAB}-nodes $\Phi_{\mathrm{s},\mathrm{L}}$ and {PPP} {NLoS} {IAB}-nodes $\Phi_{\mathrm{s},\mathrm{N}}$ with densities of $\lambda_{\mathrm{s},\mathrm{L}}(r)=\lambda_\mathrm{s}p_\mathrm{L}(r)$ and $\lambda_{\mathrm{s},\mathrm{N}}(r)=\lambda_\mathrm{s}p_\mathrm{N}(r)$, respectively. 

\subsubsection{Path Loss Model}
The path loss model utilized in \cite{7110547} was adopted in this study to measure the path loss between two nodes located at $\mathbf{x}$ and $\mathbf{y}$. The gain for a single antenna element is given by
\begin{equation}
    {L}(\mathbf{x},\mathbf{y})[\text{dB}]=\beta+10\alpha\log_{10}R_{\mathbf{x}\mathbf{y}}+\chi,
\end{equation}
where $\beta=20\log_{10}\left(\tfrac{4\pi f_\mr{c}}{c}\right)$. $f_\mr{c}$, $c$, $\alpha$, and $\chi\sim\mathcal{N}(0,\zeta^2)$ denote the carrier frequency, light speed, path loss exponent, and shadowing effect random variable, respectively. $R_{\mathbf{x}\mathbf{y}}$ represents the Euclidean distance between $\mathbf{x}$ and $\mathbf{y}$. Therefore, we have ${L}(\mathbf{x},\mathbf{y})=\tfrac{R^\alpha_{\mathbf{x}\mathbf{y}}}{10^{-0.1(\beta+\chi)}}$, where $10^{-0.1(\beta+\chi)}\sim\mr{ln}\mathcal{N}\left(-0.1\beta\mr{ln}10,\left({0.1\zeta\mr{ln}10}\right)^2\right)$ is a lognormal distributed random variable\footnote{If $Z$ is a normally distributed random variable with zero mean and variance $\sigma^2$, then $X=e^{m+nZ}$ is a lognormal random variable with parameters $\hat{\mu}=m$ and $\hat{\sigma}=n\sigma$, denoted as $X\sim\mr{ln}\mathcal{N}(\hat{\mu},\hat{\sigma}^2)$ with $\mathbb{E}\{X\}=e^{\hat{\mu}+0.5\hat{\sigma}^2}$.}. Note that the path loss exponent and Gaussian distributed shadowing effect random variable for the transmissions from gNB are denoted as $\alpha_\mr{m}$ and $\chi_\mr{m}\sim\mathcal{N}(0,\zeta_\mr{m}^2)$, respectively. For the IAB-node, we have the path loss exponent $\alpha_{\mr{s},i}$ and shadowing effect random variables $\chi_{\mr{s},i}\sim\mathcal{N}(0,\zeta_{\mr{s},i}^2)$, where $i=\mr{L}$ for the LoS IAB-node and $i=\mr{N}$ for the NLoS IAB-node.

\subsubsection{Channel Model}
Assume an OFDM single-path (i.e., flat fading) channel for all blockage scenarios, which leverages the sparsity of the FR2 band channel. A multi-path case will be left in the future. At subcarrier $k=1,2,\ldots,K$, the wideband mmWave channel from the node at $\mathbf{x}$ with $N_\mr{T}$ transmit antennas to the node at $\mathbf{y}$ with $N_\mr{R}$ receive antennas is given as \cite{7448873}
\begin{align}
\mathbf{{H}}_{\mathbf{yx}}[k]=&\sqrt{\tfrac{{N}_{\mr{T}}{N}_\mr{R}}{{L}(\mathbf{x},\mathbf{y})}}h^{\mathbf{yx}}\gamma[k]\mathbf{a}_{\mr{R}}(\bar{\theta}_\mr{R}^{\mathbf{yx}},\bar{\phi}_\mr{R}^{\mathbf{yx}})\mathbf{a}^H_{\mr{T}}(\bar{\theta}_\mr{T}^{\mathbf{yx}},\bar{\phi}_\mr{T}^{\mathbf{yx}}),\label{channel}
\end{align}
where $h^{\mathbf{yx}}$ is the small fading gain with a Nakagami-$M$ distribution \cite{NAKAGAMI19603,6015565}. $M\in\{M_\mr{m},M_{\mr{s},\mr{L}},M_{\mr{s},\mr{N}}\}\in\mathbb{N}^+$ are Nakagami parameters for the gNB, LoS IAB-node, and NLoS IAB-node, respectively. $\gamma[k]=\sum_{d=0}^{D-1}p(d-\widetilde{\tau})e^{-j\tfrac{2{\pi}kd}{K}}$ with $p(\cdot)$ is the raised-cosine pulse-shaping filter, $\widetilde{\tau}\sim\mathcal{U}[0,D]$ being the normalized path delay, and $D$ denotes the number of delay taps.

For simplicity, we leverage the virtual transmit and receive steering vectors $\mathbf{a}_{\mr{T}}(\bar{\theta}_
\mr{T}^\mathbf{yx},\bar{\phi}_
\mr{T}^\mathbf{yx})$ and $\mathbf{a}_{\mr{R}}(\bar{\theta}_
\mr{R}^\mathbf{yx},\bar{\phi}_
\mr{R}^\mathbf{yx})$ for the uniform planar array (UPA), respectively, where the azimuth $\bar{\theta}_
\mr{T}^\mathbf{yx}$/$\bar{\theta}_
\mr{R}^\mathbf{yx}$ and elevation $\bar{\phi}_
\mr{T}^\mathbf{yx}$/$\bar{\phi}_
\mr{R}^\mathbf{yx}$ virtual angles correspond to the angle of departure/arrival (AoD/AoA) \cite{1033686}. With half-wavelength spaced elements, we have the relationship between the virtual and physical azimuth (elevation) AoD/AoA as $\bar{\theta}=\pi\cos\theta\cos\phi$ ($\bar{\phi}=\pi\sin\theta\cos\phi$), where $\theta\sim\mathcal{U}[0,\pi]$ and $\phi\sim\mathcal{U}[-\tfrac{\pi}{2},\tfrac{\pi}{2}]$ are the physical azimuth and elevation AoD/AoA, respectively. By extending the model in \cite{1033686}, we quantize the virtual azimuth and elevation AoD/AoA to set $\bar{\theta}\in\left\{-\pi,-\pi+\tfrac{2\pi}{N_\mathrm{x}},\ldots,-\pi+\tfrac{2\pi (N_\mathrm{x}-1)}{N_\mathrm{x}}\right\}$ and $\bar{\phi}\in\left\{0,\tfrac{\pi}{N_\mathrm{y}},\ldots,\tfrac{\pi (N_\mathrm{y}-1)}{N_\mathrm{y}}\right\}$ with ${N_\mathrm{x}}$ and ${N_\mathrm{y}}$ being the number of antennas along $x$- and $y$-axis of the UPA. Moreover, we assume that virtual angles are uniformly distributed in their quantized sets for tractability \cite{7434598}. From the orthogonality of the virtual steering vector, we have $\mathbf{a}^H(\bar{\theta}_1,\bar{\phi}_1)\mathbf{a}(\bar{\theta}_2,\bar{\phi}_2)=1$ for $\bar{\theta}_1=\bar{\theta}_2, \bar{\phi}_1=\bar{\phi}_2$ and 0 otherwise \cite[Corollary 2]{6292865}, which gives us an ON/OFF model. Using this ON/OFF model, $\mr{SINR}$ expressions can be simplified.

Because this work aims to evaluate the performance of IBFD-IAB networks after the SI has been successfully suppressed, only the residual SI, treated as Gaussian noise, whose power is modeled as a fraction of the transmit power, will be considered. Readers are directed to \cite{Zhan2012} for a hypothetical Rician-like SI channel model and staged SIC strategy.

\section{Antenna Configuration and Beamforming Design}
\subsection{Antenna Configuration}\label{C1}
The gNB and IAB-node transmitters are adopted with subarray-based hybrid beamforming (i.e., one RF chain connects with only a portion of the antennas), as proposed in \cite{Zhan2012,7448873}. The gNB has $N_\mr{m}$ subarrays, each having $\tfrac{N_\mr{T}^\mr{m}}{N_\mr{m}}$ transmit antennas and one RF chain sending a single data stream to serve one UE or IAB-node. Similarly, the IAB-node is equipped with $N_\mr{s}$ subarrays, each with $\tfrac{N_\mr{T}^\mr{s}}{N_\mr{s}}$ transmitting antennas and a single RF chain sending one data stream to serve a UE. The IAB-node receiver is equipped with $N_\mr{R}^\mr{s}$ receiving antennas adopting analog combining to receive information from its serving gNB. The UE receiver has $N_\mr{R}^\mr{u}$ antennas with analog combining for communicating with its serving BS.

\subsection{Beamforming Design}\label{E}
According to the subarray hybrid beamforming structure given in \cite{7448873}, the RF and BB precoders at the gNBs and IAB-nodes have the form of $\mathbf{F}_{\rf}^{(\cdot)}=\mr{blkdiag}\big[\mathbf{f}_{\rf,1}^{(\cdot)},$ $\mathbf{f}_{\rf,2}^{(\cdot)},\ldots,\mathbf{f}_{\rf,N_i}^{(\cdot)}\big]\in\mathbb{C}^{N_\mr{T}^{i}\times N_i}$ and $\mathbf{F}_{\bb}^{(\cdot)}[k]=\big[\mathbf{f}_{\bb,1}^{(\cdot)}[k],$ $\mathbf{f}_{\bb,2}^{(\cdot)}[k],\ldots,\mathbf{f}_{\bb,N_i}^{(\cdot)}[k]\big]\in\mathbb{C}^{N_i\times N_i}$, respectively, where $(\cdot)$ is the location of the gNB or IAB-node; $i=\mr{m}$ represents the precoders at the gNB and $i=\mr{s}$ are those at the IAB-node. The RF combiner at the IAB-node and UE are in the form of $\mathbf{w}_{\rf}^{(\cdot)}\in\mathbb{C}^{N_\mr{R}^i\times1}$, where $i=\mr{s}$ denotes the combiner at the IAB-node and $i=\mr{u}$ is that at the UE.

RF beamformers are realized using phase shifters (PSs) with unit norm entries. According to \cite{7434598}, the optimal RF precoder/combiner that maximizes the desired signal power is to steer the beam in the direction of the AoD/AoA of the desired signal channel. Given the channel from the transmitter at $\mathbf{x}$ to the receiver at $\mathbf{y}$ in \eqref{channel}, owing to the use of subarray-based hybrid beamforming, a subset of the virtual transmit steering vector is used to design the $v$th column of the RF precoder matrix $\mathbf{F}_{\rf}^{\mathbf{x}}$, that is,
\begin{equation}
    \mathbf{f}^{\mathbf{x}}_{\rf,v}=\sqrt{N_\mr{T}^i}\left[\mathbf{a}_\mr{T}(\bar{\theta}_\mr{T}^{\mathbf{yx}},\bar{\phi}_\mr{T}^{{\mathbf{yx}}})\right]_{(v-1)\tfrac{N_\mr{T}^i}{N_i}+\left(1:\tfrac{N_\mr{T}^i}{N_i}\right),:},\label{RF}
\end{equation}
where $i=\mr{m}$ represents the RF precoder at the gNB, and $i=\mr{s}$ denotes that at the IAB-node.

Similarly, the RF combiner is given by the virtual receiver steering vector, that is,
\begin{equation}
    \mathbf{w}_{\rf}^{\mathbf{y}}=\sqrt{N_\mr{R}^i}\mathbf{a}_\mr{R}(\bar{\theta}_\mr{R}^{\mathbf{yx}},\bar{\phi}_\mr{R}^{\mathbf{yx}}),
\end{equation}
where $i=\mr{s}$ denotes the combiner at the IAB-node, and $i=\mr{u}$ denotes that at the UE.

To mitigate the intra-cell interference, the zero-forcing (ZF) BB precoder is utilized at both the gNB and IAB-node. For a transmitter at $\mathbf{x}$, given all its serving receivers at $\mathbf{y}_1,\mathbf{y}_2,\ldots,\mathbf{y}_{N_i}$ (for $i=\mr{m}$ if $\mathbf{x}$ is a gNB and  $i=\mr{s}$ if $\mathbf{x}$ is an IAB-node), the ZF BB precoder matrix can be given as
\begin{equation}
    \mathbf{F}_{\bb}^{\mathbf{x}}[k]=\widetilde{\mathbf{H}}_{\mathbf{yx}}^H[k]\left(\widetilde{\mathbf{H}}_{\mathbf{yx}}[k]\widetilde{\mathbf{H}}_{\mathbf{yx}}^H[k]\right)^{-1},
\end{equation}
where $\widetilde{\mathbf{H}}_{\mathbf{yx}}[k]=\left(\mathbf{W}_{\rf}^{\mathbf{y}}\right)^H{\hat{\mathbf{H}}}_{\mathbf{yx}}[k]\mathbf{F}_{\rf}^{\mathbf{x}}$ with $\mathbf{W}_{\rf}^{\mathbf{y}}=\mr{blkdiag}[\mathbf{w}_{\rf}^{\mathbf{y}_{1}},\mathbf{w}_{\rf}^{\mathbf{y}_{2}},\ldots,\mathbf{w}_{\rf}^{\mathbf{y}_{N_i}}]$ and $\hat{\mathbf{H}}_{\mathbf{yx}}[k]=[\mathbf{H}^T_{\mathbf{y}_{1}\mathbf{x}}[k],$ $\mathbf{H}^T_{\mathbf{y}_{2}\mathbf{x}}[k],\ldots,\mathbf{H}^T_{\mathbf{y}_{N_i}\mathbf{x}}[k]]^T$.

\section{$\mr{SINR}$ Characterization}\label{SINR}
In this section, we derive $\mr{SINR}$ expressions for the $k$th subcarrier. According to Slivnyak's theorem \cite{haenggi_2012}, given that both IAB-nodes and UEs are homogeneous PPP, we consider typical backhaul and access links without loss of generalization, that is, we derive the typical backhaul (access) link $\mr{SINR}$ by assuming that the typical IAB-node (UE) is located at the origin $\hat{\mathbf{0}}$ ($\mathbf{0}$). Note that ${\mathbf{0}}$ and $\hat{\mathbf{0}}$ represent the origins of the same coordinate system.

Assume that we know the channel state information perfectly for all channels, and that all signals are uncorrelated. Let the average total transmit power at the gNB (IAB-node) across all subcarriers be $P_\mr{m}$ $\left(P_\mr{s}\right)$. With equal power allocation, each data stream at subcarrier $k$ transmitted from the gNB (IAB-node) has the power of $\tfrac{P_\mr{m}}{KN_\mr{m}}$ $\left(\tfrac{P_\mr{s}}{KN_\mr{s}}\right)$. Moreover, as the ZF precoder is applied, the intra-cell interference can be canceled out; thus, is ignored in the following equations. In addition, the following assumption was made simplify the analysis.

\textit{Assumption 1: The channel model in Eq.~\eqref{channel} is a function of $k$, where $k$ appears only in $\gamma[k]$. Because our work is based on a probabilistic perspective, for tractability, we assume the normalized path delay to be $\widetilde{\tau}=\tfrac{D}{2}$, which is the mean of the random variable $\widetilde{\tau}$. Given that the channel is flat fading, all subcarriers experience the same channel conditions. Thus, we have $\mathbb{E}\{|\gamma[k]|^2\}=1\forall k$, which makes the channel independent of $k$ \cite{tong}.}

\subsubsection{Backhaul link}
Assuming that the gNB at $\mathbf{x}_\mr{m}$ uses its $v$th subarray to provide backhaul communication with the typical IAB-node at $\hat{\mathbf{0}}$, the $\mr{SINR}$ of the typical backhaul link at the $k$th subcarrier is written as
\begin{subequations}
\begin{align}
    &\mr{SINR}_{\hat{\mathbf{0}}\mathbf{x}_\mr{m}}[k]=\tfrac{G_\mr{b}[k]}{I_\mr{m,b}[k]+I_\mr{s,b}[k]+\underbrace{\tfrac{\eta P_\mr{s}}{K}}_{\text{RSI}}+\mathbb{E}\left\{\left|\left(\mathbf{w}_{\rf}^{\hat{\mathbf{0}}}\right)^H\mathbf{n}_\mr{b}[k]\right|^2\right\}+\mathbb{E}\left\{\left|e_\mr{b}[k]\right|^2\right\}},\label{sinrb}\\
    &G_\mr{b}[k]=\frac{P_\mr{m}}{KN_\mr{m}}\left|\left(\mathbf{w}_{\rf}^{\hat{\mathbf{0}}}\right)^H\mathbf{H}_{\hat{\mathbf{0}}\mathbf{x}_\mr{m}}[k]\mathbf{F}_{\rf}^{\mathbf{x}_\mr{m}}\mathbf{f}_{\bb,v}^{\mathbf{x}_\mr{m}}[k]\right|^2\notag\\&\qquad\;\approx\frac{P_\mr{m}N_\mr{T}^\mr{m}\left(N_\mr{R}^\mr{s}\right)^2\widetilde{h}^{\hat{\mathbf{0}}\mathbf{x}_\mr{m}}p_\mr{ZF}(N_\mr{T}^\mr{m},N_\mr{m})}{KN^2_\mr{m}R_{\mathbf{x}_\mr{m}\hat{\mathbf{0}}}^{\alpha_\mr{m}}},\label{gb}\\
    &I_\mr{m,b}[k]=\frac{P_\mr{m}}{KN_\mr{m}}\left\lVert\sum\limits_{{\omega}\in\atop\Phi_\mr{m}\backslash {\mathbf{x}_\mr{m}}}\left(\mathbf{w}_{\rf}^{\hat{\mathbf{0}}}\right)^H\mathbf{H}_{\hat{\mathbf{0}}{\omega}}[k]\mathbf{F}_{\rf}^{{\omega}}\mathbf{F}_{\bb}^{{\omega}}[k]\right\rVert^2\notag\\&\qquad\;\;\;\approx\sum\limits_{\omega\in\atop\Phi_\mr{m}\backslash {\mathbf{x}_\mr{m}}}\frac{P_\mr{m}N^\mr{m}_\mr{T}\left(N^{s}_\mr{R}\right)^2\widetilde{h}^{\hat{\mathbf{0}}\omega}p_\mr{BF}(N^\mr{m}_\mr{T},N^\mr{s}_\mr{R},N_\mr{m})}{KN_\mr{m}R_{\omega\hat{\mathbf{0}}}^{\alpha_\mr{m}}},\label{im}\\
    &I_\mr{s,b}[k]=\sum\limits_{i\in\{\mr{L},\mr{N}\}}I_\mr{s,b}^i[k]\notag\\&\qquad\;=\frac{P_\mr{s}}{KN_\mr{s}}\left\lVert\sum\limits_{i\in\{\mr{L},\mr{N}\}}\sum\limits_{\hat{{\omega}}\in\Phi_{\mr{s},i}\backslash {\hat{\mathbf{0}}}}\left(\mathbf{w}_{\rf}^{\hat{\mathbf{0}}}\right)^H\mathbf{H}_{\hat{\mathbf{0}}\hat{\omega}}[k]\mathbf{F}_{\rf}^{\hat{\omega}}\mathbf{F}_{\bb}^{\hat{\omega}}[k]\right\rVert^2\notag\\&\qquad\;\approx\sum\limits_{i\in\{\mr{L},\mr{N}\}}\sum\limits_{\hat{\omega}\in\Phi_{\mr{s},i}\backslash {\hat{\mathbf{0}}}}\frac{P_\mr{s}N^\mr{s}_\mr{T}\left(N^\mr{s}_\mr{R}\right)^2\widetilde{h}^{\hat{\mathbf{0}}\hat{\omega}}p_\mr{BF}(N^\mr{s}_\mr{T},N^\mr{s}_\mr{R},N_\mr{s})}{KN_\mr{s}R_{\hat{\omega}\hat{\mathbf{0}}}^{\alpha_{\mr{s},i}}},\label{is}
\end{align}
\end{subequations}
where in \eqref{sinrb}, $G_\mr{b}[k]$, $I_\mr{m,b}[k]$, and $I_\mr{s,b}[k]$ denote the power of the desired signal, the power of the interference from non-target gNBs, and that of the interference from IAB-nodes at subcarrier $k$, respectively; $0<\eta<1$ is the RSI controlling factor, which controls the amount of RSI; $\mathbb{E}\left\{\left|\left(\mathbf{w}_{\rf}^{\hat{\mathbf{0}}}\right)^H\mathbf{n}_\mr{b}[k]\right|^2\right\}=\sigma_\mr{n}^2N^\mr{s}_\mr{R}$ is the Gaussian noise power with $\mathbf{n}_\mr{b}[k]\sim\mathcal{CN}(\mathbf{0},\sigma_\mr{n}^2\mathbf{I}_{N^\mr{s}_\mr{R}})$; $\mathbb{E}\left\{\left|e_\mr{b}[k]\right|^2\right\}$ represents the quantization noise power owing to $q$ bits ADC uniform quantization. In \eqref{gb}-\eqref{is}, $\widetilde{h}^{\hat{\mathbf{0}}\mathbf{x}_\mr{m}}=10^{-0.1(\beta+\chi_\mr{m})}\left|h^{\hat{\mathbf{0}}\mathbf{x}_\mr{m}}\right|^2$, $\widetilde{h}^{\hat{\mathbf{0}}\omega}=10^{-0.1(\beta+\chi_\mr{m})}\left|h^{\hat{\mathbf{0}}\omega}\right|^2$, and $\widetilde{h}^{\hat{\mathbf{0}}\hat{\omega}}=10^{-0.1(\beta+\chi_{\mr{s},i})}\left|h^{\hat{\mathbf{0}}\hat{\omega}}\right|^2$ follow the composite GL distribution (see Appendix-\ref{GL}) because $\left|h^{(\cdot)}\right|^2$ is a Gamma distributed random variable\footnote{If $X$ is a Nakagami-$M$ distributed random variable, then $X^2\sim \mr{Gamma}(M,1/M)$ is a unit mean Gamma-distributed random variable.}; $p_\mr{ZF}(N_\mr{T}^\mr{m},N_\mr{m})$ is the ZF penalty, which is a Bernoulli random variable with success probability $(1-\tfrac{1}{N_\mr{T}^\mr{m}})^{N_\mr{m}-1}$ (see Remark 1); $p_\mr{BF}(N^\mr{m}_\mr{T},N^\mr{s}_\mr{R},N_\mr{m})$ and $p_\mr{BF}(N^\mr{s}_\mr{T},N^\mr{s}_\mr{R},N_\mr{s})$ are beamforming gains that take into account the sidelobe gains (see Proposition 1) because using the ON/OFF model given by the orthogonality of the steering vector can result in underestimation of the inter-cell interference as it gives zero beamforming gain when the azimuth and/or elevation angles are not the same. The approximation in \eqref{gb} is owing to the ZF penalty. The approximation in \eqref{im} is obtained by assuming $\mathbf{F}_{\bb}^{{\omega}}[k]=\sqrt{\tfrac{N_\mr{T}^\mr{m}}{N_\mr{m}}}\mathbf{I}_{N_\mr{m}}$, where $\sqrt{\tfrac{N_\mr{T}^\mr{m}}{N_\mr{m}}}$ ensures $\left\|\mathbf{F}_{\rf}^{\omega}\mathbf{F}_{\bb}^{{\omega}}\right\|_F^2=N_\mr{m}$. For a large number of antennas, the ZF penalty and the effect of the ZF on the inter-cell interference link can be neglected, as evaluated in \cite{7434599}, and \eqref{is} follows a similar approximation.

\textit{Remark 1: For notation simplicity, we assume that the transmitter at $\mathbf{x}$ uses its $1$st subarray to communicate with the typical receiver at $\mathbf{y}$. By adopting the orthogonality of the steering vector, given the transmit steering vector $\mathbf{a}_{\mr{T}}(\bar{\theta}_
\mr{T}^{\mathbf{yx}},\bar{\phi}_
\mr{T}^{\mathbf{yx}})$, we have $\mathbf{a}_{\mr{T}}(\bar{\theta}_
\mr{T}^{\mathbf{yx}},\bar{\phi}_
\mr{T}^{\mathbf{yx}})\mathbf{F}_{\rf}^{\mathbf{x}}=\left[\tfrac{\sqrt{N_\mr{T}^i}}{N_i},0,\ldots,0\right]\in\mathbb{R}^{1\times N_i}$ only occurs when the AoD for the typical link is different from that for others served by the transmitter at $\mathbf{x}$, which has a probability of $\left(1-\tfrac{1}{N_\mr{T}^i}\right)^{N_i-1}$ for $p_\mr{ZF}(N_\mr{T}^i,N_i)=1$. Accordingly, we have $\mathbf{f}_{\bb,1}^{\mathbf{x}}[k]=\left[\sqrt{\tfrac{N_i}{N_\mr{T}^i}},0,\ldots,0\right]^T$ satisfies $\left\|\mathbf{F}_{\rf}^{\mathbf{x}}\mathbf{f}_{\bb,1}^{\mathbf{x}}[k]\right\|_F^2=1$. For a large number of antennas, other cases occur with very low probability, that is, $1-\left(1-\tfrac{1}{N_\mr{T}^i}\right)^{N_i-1}$, for tractability, the signal power is approximated to be 0 by setting $p_\mr{ZF}(N_\mr{T}^i,N_i)=0$. Note that $i=\mr{m}$ denotes the transmitter at the gNB and $i=\mr{s}$ denotes that at the IAB-node \cite{7434598}.}

\textit{Proposition 1: The beamforming gain (including sidelobe gains) of the inter-cell interference links is given by $p_\mr{BF}(N_\mr{T}^i,N_\mr{R}^j,N_i)$, which can be chosen from one of the values with the corresponding probability of occurrence in \eqref{BFpenlty} shown on top of the next page,
\newcounter{myempeqncnt}
\begin{figure*}[htbp]
\normalsize
\setcounter{equation}{8}
\begin{align}
&p_\mathrm{BF}(N_\mathrm{T}^i,N_\mathrm{R}^j,N_i)=\left\{
\begin{aligned}
&p+qg_{\mr{T},i}^2+(N_i-p-q){\hat{g}_{\mr{T},i}^2}\qquad\qquad\text{w.p.}\;\tfrac{1}{N_\mr{R}^j}C_{N_i}^pC_{N_i-p}^q\left(\tfrac{1}{N_\mr{T}^i}\right)^p\left(1-\tfrac{1}{N_\mr{T,x}^i}-\tfrac{1}{N_\mr{T,y}^i}+\tfrac{1}{N_\mr{T}^i}\right)^q\\&\;\qquad\;\quad\qquad\qquad\qquad\qquad\qquad\qquad\qquad\times\left(\tfrac{1}{N_\mr{T,x}^i}+\tfrac{1}{N_\mr{T,y}^i}-\tfrac{2}{N_\mr{T}^i}\right)^{N_i-p-q}\\
&g_{\mr{R},j}^2\left[p+qg_{\mr{T},i}^2+(N_i-p-q){\hat{g}_{\mr{T},i}}^2\right]\quad\;\text{w.p.}\;\left(1-\tfrac{1}{N_\mr{R,x}^j}-\tfrac{1}{N_\mr{R,y}^j}+\tfrac{1}{N_\mr{R}^j}\right)C_{N_i}^pC_{N_i-p}^q\left(\tfrac{1}{N_\mr{T}^i}\right)^p\\&\;\qquad\;\quad\qquad\qquad\qquad\qquad\qquad\qquad\qquad\times\left(1-\tfrac{1}{N_\mr{T,x}^i}-\tfrac{1}{N_\mr{T,y}^i}+\tfrac{1}{N_\mr{T}^i}\right)^q\left(\tfrac{1}{N_\mr{T,x}^i}+\tfrac{1}{N_\mr{T,y}^i}-\tfrac{2}{N_\mr{T}^i}\right)^{N_i-p-q}\\
&\hat{g}_{\mr{R},j}^2\left[p+qg_{\mr{T},i}^2+(N_i-p-q){\hat{g}_{\mr{T},i}}^2\right]\quad\;\text{w.p.}\;\left(\tfrac{1}{N_\mr{R,x}^j}+\tfrac{1}{N_\mr{R,y}^j}-\tfrac{2}{N_\mr{R}^j}\right)C_{N_i}^pC_{N_i-p}^q\left(\tfrac{1}{N_\mr{T}^i}\right)^p\\&\;\qquad\;\quad\qquad\qquad\qquad\qquad\qquad\qquad\qquad\times\left(1-\tfrac{1}{N_\mr{T,x}^i}-\tfrac{1}{N_\mr{T,y}^i}+\tfrac{1}{N_\mr{T}^i}\right)^q\left(\tfrac{1}{N_\mr{T,x}^i}+\tfrac{1}{N_\mr{T,y}^i}-\tfrac{2}{N_\mr{T}^i}\right)^{N_i-p-q},
\end{aligned}
\right.\label{BFpenlty}
\end{align}
\hrulefill
\setcounter{equation}{\value{myempeqncnt}}
\end{figure*}
for $p,q\in\mathbb{N}$, $p+q\leq {N_i}$, where $N_\mr{T,x}^i$ and $N_\mr{T,y}^i$ ($N_\mr{R,x}^j$ and $N_\mr{R,y}^j$) denote the number of antennas along $x$- and $y$-axis of the transmit (receive) UPA. $g_{\mr{T},i}$ and $\hat{g}_{\mr{T},i}$ ($g_{\mr{R},j}$ and $\hat{g}_{\mr{R},j}$) are the sidelobe gains at the transmitter (receiver). $i=\mr{m}$, $j=\mr{s}$; $i=\mr{s}$, $j=\mr{s}$; $i=\mr{m}$, $j=\mr{u}$; and $i=\mr{s}$, $j=\mr{u}$ denote the beamforming gain of the interference from gNBs to the typical IAB-node, IAB-nodes to the typical IAB-node, gNBs to the typical UE, and IAB-nodes to the typical UE, respectively.}
\begin{IEEEproof}
Assume squared UPA, the inner product between transmit steering vectors is $\mathbf{a}_{\mr{T},i}^H(\bar{\theta}_1,\bar{\phi}_1)\mathbf{a}_{\mr{T},i}(\bar{\theta}_2,\bar{\phi}_2)=$
\setcounter{equation}{9}
{\begin{align}
  \left\{
\begin{aligned}
\quad &1 \quad\;\;\;\;\;\qquad\qquad\qquad\qquad\;\text{if}\;\bar{\theta}_1=\bar{\theta}_2, \bar{\phi}_1=\bar{\phi}_2\\
&g_{\mr{T},i}=\tfrac{1}{\sin^2(0.244)N_\mr{T,x}^iN_\mr{T,y}^i} \quad\text{if}\;\bar{\theta}_1\neq\bar{\theta}_2, \bar{\phi}_1\neq\bar{\phi}_2\\
&\hat{g}_{\mr{T},i}=\tfrac{1}{\sin(0.244)N_\mr{T,x}^i} \quad\qquad\text{otherwise},
\end{aligned}
\right.
\end{align}
where the sidelobe gains $g_{\mr{T},i}$ and $\hat{g}_{\mr{T},i}$ are derived according to \cite{antenna,7434599}.}
Similarly, we can model the inner product between the receiver steering vectors with sidelobe gains $g_{\mr{R},j}$ and $\hat{g}_{\mr{R},j}$. Recall that the virtual angles are uniformly distributed in their quantization sets, and according to the beamformer design in Section~\ref{E}, this proposition can be easily proven.
\end{IEEEproof}

According to \cite{roberts2020hybrid}, the ADC quantization noise power can be approximated as
\begin{equation}
    \mathbb{E}\left\{\left|e_\mr{b}[k]\right|^2\right\}\approx\frac{1}{K}\sum_{k'=0}^{K-1}\tfrac{G_\mr{b}[k']+\eta_\mr{dig}\tfrac{\eta P_\mr{s}}{K}+I_\mr{m,b}[k']+I_\mr{s,b}[k']+\sigma_\mr{n}^2N^\mr{s}_\mr{R}}{1.5\cdot2^{2q}},\label{quan}
\end{equation} 
where $\eta_\mr{dig}>1$ is the amount of digital cancellation, introduced to reproduce the RSI power before the ADC (i.e., after analog SIC) because $\tfrac{\eta P_\mr{s}}{K}$ is the RSI power after the ADC (i.e., after digital SIC). $q$ denotes the ADC quantization bits. Owing to Assumption 1, the value of the numerator of \eqref{quan} is the same for all subcarriers. Thus, \eqref{quan} can be simplified as
\begin{equation}
    \mathbb{E}\left\{\left|e_\mr{b}[k]\right|^2\right\}\approx\frac{G_\mr{b}[k]+\eta_\mr{dig}\tfrac{\eta P_\mr{s}}{K}+I_\mr{m,b}[k]+I_\mr{s,b}[k]+\sigma_\mr{n}^2N^\mr{s}_\mr{R}}{1.5\cdot2^{2q}}.\label{quan1}
\end{equation}

\subsubsection{Access link}
Next, assuming that gNB at $\mathbf{x}_\mr{m}$ uses its $u$th subarray to communicate with the typical UE at $\mathbf{0}$, we derive the $\mr{SINR}$ expression of the gNB-associated typical access link at subcarrier $k$, which is cast as
\begin{subequations}
\begin{align}
    &\mr{SINR}_{{\mathbf{0}}\mathbf{x}_\mr{m}}[k]=\tfrac{G_\mr{a}[k]}{I_\mr{m,a}[k]+I_\mr{s,a}[k]+\mathbb{E}\left\{\left|\left(\mathbf{w}_{\rf}^{{\mathbf{0}}}\right)^H\mathbf{n}_\mr{a,m}[k]\right|^2\right\}+\mathbb{E}\left\{\left|e_\mr{a,m}[k]\right|^2\right\}},\label{sinra}\\
    &G_\mr{a}[k]=\frac{P_\mr{m}}{KN_\mr{m}}\left|\left(\mathbf{w}_{\rf}^{{\mathbf{0}}}\right)^H\mathbf{H}_{{\mathbf{0}}\mathbf{x}_\mr{m}}[k]\mathbf{F}_{\rf}^{\mathbf{x}_\mr{m}}\mathbf{f}_{\bb,u}^{\mathbf{x}_\mr{m}}[k]\right|^2\notag\\&\qquad\;\approx\frac{P_\mr{m}N_\mr{T}^\mr{m}\left(N_\mr{R}^\mr{u}\right)^2\widetilde{h}^{{\mathbf{0}}\mathbf{x}_\mr{m}}p_\mr{ZF}(N_\mr{T}^\mr{m},N_\mr{m})}{KN^2_\mr{m}R_{\mathbf{x}_\mr{m}{\mathbf{0}}}^{\alpha_\mr{m}}},\label{gb1}\\
    &I_\mr{m,a}[k]=\frac{P_\mr{m}}{KN_\mr{m}}\left\lVert\sum\limits_{{\omega}\in\Phi_\mr{m}\backslash {\mathbf{x}_\mr{m}}}\left(\mathbf{w}_{\rf}^{{\mathbf{0}}}\right)^H\mathbf{H}_{{\mathbf{0}}{\omega}}[k]\mathbf{F}_{\rf}^{{\omega}}\mathbf{F}_{\bb}^{{\omega}}[k]\right\rVert^2\notag\\&\qquad\;\;\;\approx\sum\limits_{\omega\in\Phi_\mr{m}\backslash {\mathbf{x}_\mr{m}}}\frac{P_\mr{m}N^\mr{m}_\mr{T}\left(N^{u}_\mr{R}\right)^2\widetilde{h}^{{\mathbf{0}}\omega}p_\mr{BF}(N^\mr{m}_\mr{T},N^\mr{u}_\mr{R},N_\mr{m})}{KN_\mr{m}R_{\omega{\mathbf{0}}}^{\alpha_\mr{m}}},\label{im1}\\
    &I_\mr{s,a}[k]=\sum\limits_{i\in\{\mr{L},\mr{N}\}}I_\mr{s,a}^i[k]\notag\\&\qquad\;=\frac{P_\mr{s}}{KN_\mr{s}}\left\lVert\sum\limits_{i\in\{\mr{L},\mr{N}\}}\sum\limits_{\hat{{\omega}}\in\Phi_{\mr{s},i}\backslash {{\mathbf{0}}}}\left(\mathbf{w}_{\rf}^{{\mathbf{0}}}\right)^H\mathbf{H}_{{\mathbf{0}}\hat{\omega}}[k]\mathbf{F}_{\rf}^{\hat{\omega}}\mathbf{F}_{\bb}^{\hat{\omega}}[k]\right\rVert^2\notag\\&\qquad\;\approx\sum\limits_{i\in\{\mr{L},\mr{N}\}}\sum\limits_{\hat{\omega}\in\Phi_{\mr{s},i}\backslash {{\mathbf{0}}}}\frac{P_\mr{s}N^\mr{s}_\mr{T}\left(N^\mr{u}_\mr{R}\right)^2\widetilde{h}^{{\mathbf{0}}\hat{\omega}}p_\mr{BF}(N^\mr{s}_\mr{T},N^\mr{u}_\mr{R},N_\mr{s})}{KN_\mr{s}R_{\hat{\omega}{\mathbf{0}}}^{\alpha_{\mr{s},i}}},\label{is1}
\end{align}
\end{subequations}
where in \eqref{sinra}, $G_\mr{a,m}[k]$, $I_\mr{m,a}[k]$, and $I_\mr{s,a}[k]$ denote the power of the desired signal, that of the interference from non-target gNBs, and that of the interference from IAB-nodes at subcarrier $k$, respectively; $\mathbb{E}\left\{\left|\left(\mathbf{w}_{\rf}^{{\mathbf{0}}}\right)^H\mathbf{n}_\mr{a,m}[k]\right|^2\right\}=\sigma_\mr{n}^2N^\mr{u}_\mr{R}$ is the Gaussian noise power with $\mathbf{n}_\mr{a,m}[k]\sim\mathcal{CN}(\mathbf{0},\sigma_\mr{n}^2\mathbf{I}_{N^\mr{u}_\mr{R}})$; $\mathbb{E}\left\{\left|e_\mr{a,m}[k]\right|^2\right\}$ represents the ADC quantization noise power with a definition similar to that in \eqref{quan1}. In \eqref{gb1}-\eqref{is1}, $\widetilde{h}^{\mathbf{0}\mathbf{x}_\mr{m}}=10^{-0.1(\beta+\chi_\mr{m})}\left|h^{\mathbf{0}\mathbf{x}_\mr{m}}\right|^2$, $\widetilde{h}^{\mathbf{0}\omega}=10^{-0.1(\beta+\chi_\mr{m})}\left|h^{\mathbf{0}\omega}\right|^2$, and $\widetilde{h}^{\mathbf{0}\hat{\omega}}=10^{-0.1(\beta+\chi_{\mr{s},i})}\left|h^{\mathbf{0}\hat{\omega}}\right|^2$ follows from the composite GL distribution; $p_\mr{ZF}(N_\mr{T}^\mr{m},N_\mr{m})$ is the ZF penalty following Remark 1; $p_\mr{BF}(N_\mr{T}^\mr{m},N_\mr{R}^\mr{u},N_\mr{m})$ and $p_\mr{BF}(N_\mr{T}^\mr{s},N_\mr{R}^\mr{u},N_\mr{s})$ are beamforming gains according to Proposition 1.

Finally, assuming the IAB-node at $\mathbf{x}_\mr{s}^j$ with $j\in\{\mr{L},\mr{N}\}$ uses its $v$th subcarrier to provide communications to the typical UE at $\mathbf{0}$, the $\mr{SINR}$ expression of the IAB-node associated typical access link at the subcarrier $k$ is expressed as
\begin{subequations}
\begin{align}
    &\mr{SINR}_{\mathbf{0}\mathbf{x}_\mr{s}^j}[k]=\tfrac{G_\mr{a,s}[k]}{\hat{I}_\mr{m,a}[k]+\hat{I}_\mr{s,a}[k]+\mathbb{E}\left\{\left|\left(\mathbf{w}_{\rf}^{\mathbf{0}}\right)^H\mathbf{n}_\mr{a,s}[k]\right|^2\right\}+\mathbb{E}\left\{\left|e_\mr{a,s}[k]\right|^2\right\}}\label{sinraa}\\
    &G_\mr{a,s}[k]=\frac{P_\mr{s}}{KN_\mr{s}}\left|\left(\mathbf{w}_{\rf}^{\mathbf{0}}\right)^H\mathbf{H}_{\mathbf{0}\mathbf{x}_\mr{s}^j}[k]\mathbf{F}_{\rf}^{\mathbf{x}_\mr{s}^j}\mathbf{f}_{\bb,v}^{\mathbf{x}_\mr{s}^j}\right|^2\notag\\&\quad\quad\;\;\;\approx\frac{P_\mr{s}N_\mr{T}^\mr{s}\left(N_\mr{R}^\mr{u}\right)^2\widetilde{h}^{\mathbf{0}\mathbf{x}_\mr{s}^j}p_\mr{ZF}(N^\mr{s}_\mr{T},N_\mr{s})}{KN^2_\mr{s}R_{\mathbf{x}_\mr{s}^j\mathbf{0}}^{\alpha_{\mr{s},j}}},\label{gs}\\
    &\hat{I}_\mr{m,a}[k]=\frac{P_\mr{m}}{KN_\mr{m}}\left\lVert\sum\limits_{\omega\in\Phi_\mr{m}}\left(\mathbf{w}_{\rf}^{\mathbf{0}}\right)^H\mathbf{H}_{\mathbf{0}\omega}[k]\mathbf{F}_{\rf}^{\omega}\mathbf{F}_{\bb}^{\omega}\right\rVert^2\notag\\&\quad\quad\;\;\;\approx\sum\limits_{\omega\in\Phi_\mr{m}}\frac{P_\mr{m}N_\mr{T}^\mr{m}\left(N_\mr{R}^\mr{u}\right)^2\widetilde{h}^{\mathbf{0}\omega}p_\mr{BF}(N_\mr{T}^\mr{m},N_\mr{R}^\mr{u},N_\mr{m})}{KN_\mr{m}R_{\omega\mathbf{0}}^{\alpha_\mr{m}}},\\
    &\hat{I}_\mr{s,a}[k]=\sum\limits_{i\in\{\mr{L},\mr{N}\}}\hat{I}_\mr{s,a}^i[k]\notag\\&\quad\quad\;\;=\frac{P_\mr{s}}{KN_\mr{s}}\left\lVert\sum\limits_{i\in\{\mr{L},\mr{N}\}}\sum\limits_{\hat{\omega}\in\Phi_{\mr{s},i}\backslash{\mathbf{x}_\mr{s}^j}} \left(\mathbf{w}_{\rf}^{\mathbf{0}}\right)^H\mathbf{H}_{\mathbf{0}\hat{\omega}}[k]\mathbf{F}_{\rf}^{\hat{\omega}}\mathbf{F}_{\bb}^{\hat{\omega}}\right\rVert^2\notag\\&\quad\quad\;\;\approx\sum\limits_{i\in\{\mr{L},\mr{N}\}}\sum\limits_{\hat{\omega}\in\Phi_{\mr{s},i}\backslash {\mathbf{x}_\mr{s}^j}}\frac{P_\mr{s}N_\mr{T}^\mr{s}\left(N_\mr{R}^\mr{u}\right)^2\widetilde{h}^{\mathbf{0}\hat{\omega}}p_\mr{BF}(N_\mr{T}^\mr{s},N_\mr{R}^\mr{u},N_\mr{s})}{KN_\mr{s}R_{\hat{\omega}\mathbf{0}}^{\alpha_{\mr{s},i}}},\label{asinr2}
\end{align}
\end{subequations}
where in \eqref{sinraa}, $G_\mr{a,s}[k]$, $\hat{I}_\mr{m,a}[k]$, and $\hat{I}_\mr{s,a}[k]$ denote the power of the desired signal, that of the interference from gNBs, and that of the interference from non-target IAB-nodes at subcarrier $k$, respectively; $\mathbb{E}\left\{\left|\left(\mathbf{w}_{\rf}^{\mathbf{0}}\right)^H\mathbf{n}_\mr{a,s}[k]\right|^2\right\}=\sigma_\mr{n}^2N^\mr{u}_\mr{R}$ is the Gaussian noise power with $\mathbf{n}_\mr{a,s}[k]\sim\mathcal{CN}(\mathbf{0},\sigma_\mr{n}^2\mathbf{I}_{N^\mr{u}_\mr{R}})$; $\mathbb{E}\left\{\left|e_\mr{a,s}[k]\right|^2\right\}$ represents the ADC quantization noise power similar to \eqref{quan1}. In \eqref{gs}-\eqref{asinr2}, $\widetilde{h}^{\mathbf{0}\mathbf{x}_\mr{s}^j}=10^{-0.1(\beta+\chi_{\mr{s},j})}\left|h^{\mathbf{0}\mathbf{x}_\mr{s}^j}\right|^2$, $\widetilde{h}^{\mathbf{0}\omega}=10^{-0.1(\beta+\chi_\mr{m})}\left|h^{\mathbf{0}\omega}\right|^2$, and $\widetilde{h}^{\mathbf{0}\hat{\omega}}=10^{-0.1(\beta+\chi_{\mr{s},i})}\left|h^{\mathbf{0}\hat{\omega}}\right|^2$ are composite GL distributed random variables; $p_\mr{ZF}(N^\mr{s}_\mr{T},N_\mr{s})$ is the ZF penalty following Remark 1; $p_\mr{BF}(N_\mr{T}^\mr{m},N_\mr{R}^\mr{u},N_\mr{m})$ and $p_\mr{BF}(N_\mr{T}^\mr{s},N_\mr{R}^\mr{u},N_\mr{s})$ are beamforming gains following Proposition 1.

\section{Performance Analysis}
For a standard {OFDM} system, owing to different channel gains occurring on different subcarriers, the effective $\mr{SINR}$, whose statistics can be obtained by Lognormal or Gaussian approximations, is used to derive the performance metrics \cite{5285161,6213041}, that is
\begin{align}
    \mathrm{SINR}_{eff}=\widetilde{\vartheta}\mathrm{ln}\left(\frac{1}{K}\sum_{k=0}^Ke^{\frac{\mathrm{SINR}[k]}{\widetilde{\vartheta}}}\right),
\end{align}
where $\widetilde{\vartheta}$ is a parameter that depends on the coding scheme and number of coded bits in a block. However, in this work, thanks to Assumption 1, the value of $\mr{SINR}$ is the same for all subcarriers, which means $\mathrm{SINR}[k]\;\forall \;k$ equals to the effective $\mr{SINR}$. Therefore, $\mathrm{SINR}[k]$ was utilized for the analysis, and the subcarrier index was omitted for simplicity.

\subsection{Association Probabilities}\label{3-A}
This subsection introduces the association probabilities; that is, the probabilities of a typical UE served by a gNB and an IAB-node, based on the maximum \textit{long-term averaged biased-received-desired-signal power} (i.e., $\mathbb{E}\{\text{received-desired-signal power}\}\times\text{bias factor}\;T_i$, for $i=\mr{m}$ if the desired signal comes from the gNB and $i=\mr{s}$ if it comes from the IAB-node) criteria \cite{6287527}. The bias factor plays the role of offloading the traffic from gNB to IAB-node (or from IAB-node to gNB) to satisfy network-wide performance requirement \cite{6497002}. The $\mr{SINR}$ coverage could be changed by tuning the bias ratio $\tfrac{T_\mr{s}}{T_\mr{m}}$, as shown in Section~\ref{simulation}.

To derive the association probabilities, in the following lemmas, we first show the probability density function (PDF) and cumulative distribution function (CDF) of the contact distance (i.e., the distance of a typical point in a point process to its nearest point in an independent point process) for three different cases:
\begin{itemize}
\item[(i)] The typical UE in PPP to its nearest LoS IAB-node in an independent PPP.
\item[(ii)] The typical UE in PPP to its nearest NLoS IAB-node in an independent PPP.
\item[(iii)] The typical UE in PPP to its nearest gNB in an independent MHCPP.
\end{itemize}

\textit{Lemma 1: The CDF of the contact distance from the PPP typical UE to the PPP LoS/NLoS IAB-node with density $\lambda_{\mr{s},\mr{L}}(x)/\lambda_{\mr{s},\mr{N}}(x)$} is cast as
\begin{align}
    F_{R_{\mathbf{x}_\mr{s}^i\mathbf{0}}}(r_{\mr{s},i})=1-e^{-\int_0^{r_{\mr{s},i}}2\pi x\lambda_{\mr{s},i}(x)\mr{d}x}.
\end{align}
The corresponding PDF is written as
\begin{align}
    f_{R_{\mathbf{x}_\mr{s}^i\mathbf{0}}}(r_{\mr{s},i})=2\pi\lambda_{\mr{s},i}(r_{\mr{s},i})e^{-\int_0^{r_{\mr{s},i}}2\pi x\lambda_{\mr{s},i}(x)\mr{d}x},
\end{align}
where $r_{\mr{s},i}>0$ denotes the distance from the PPP typical UE to the PPP LoS IAB-node for $i=\mr{L}$; and that to the PPP NLoS IAB-node for $i=\mr{N}$.
\begin{IEEEproof}
The similar proof can be found in \cite{7511676}.
\end{IEEEproof}

Regarding the PDF and CDF of the PPP to MHCPP contact distance, it is impossible to obtain exact expressions because of the correlations among the MHCPP points. Fortunately, the authors in \cite{8060551} provided a piece-wise model, as shown in the following lemma, which has been verified to provide a very close solution compared with the empirical data. 

\textit{Lemma 2: The CDF of the contact distance from the PPP typical UE to the MHCPP gNB with density $\lambda_\mr{m}$ and hard-core distance $\xi$} is approximated as
\begin{align}
F_{R_{\mathbf{x}_\mr{m}\mathbf{0}}}(r_\mr{m})=\left\{
\begin{aligned}
F_{R_{\mathbf{x}_\mr{m}\mathbf{0}},1}(r_\mr{m})&=\pi\lambda_\mr{m}r_\mr{m}^2 \quad\;\;{0\leq r_\mr{m}\leq\tfrac{\xi}{2}},\\
F_{R_{\mathbf{x}_\mr{m}\mathbf{0}},2}(r_\mr{m})&=1-\frac{4-\pi\lambda_\mr{m} \xi^2}{4}\\&\times e^{\tfrac{2\pi\lambda_\mr{m}\xi^2\left[1-\left(\tfrac{2r_\mr{m}}{\xi}\right)^\varrho\right]}{\varrho\left(4-\pi\lambda_\mr{m}\xi^2\right)}}  \\&\qquad\qquad\qquad\;\;\quad{r_\mr{m}>\tfrac{\xi}{2}}.\end{aligned}
\right.\label{MHCPPCDF}
\end{align}
Taking the first derivative with respect to $r_\mr{m}$, we obtain the PDF of the contact distance as
\begin{align}
f_{R_{\mathbf{x}_\mr{m}\mathbf{0}}}(r_\mr{m})=\left\{
\begin{aligned}
f_{R_{\mathbf{x}_\mr{m}\mathbf{0}},1}(r_\mr{m})&=2\pi\lambda_\mr{m} r_\mr{m} \quad\qquad\;{0\leq r_\mr{m}\leq\tfrac{\xi}{2}},\\
f_{R_{\mathbf{x}_\mr{m}\mathbf{0}},2}(r_\mr{m})&=2\pi\lambda_\mr{m} r_\mr{m}\left(\tfrac{2r}{\xi}\right)^{\varrho-2}\\&\times e{\tfrac{2\pi\lambda_\mr{m}\xi^2\left[1-\left(\tfrac{2r_\mr{m}}{\xi}\right)^\varrho\right]}{\varrho\left(4-\pi\lambda_\mr{m}\xi^2\right)}} \;\;{r_\mr{m}>\tfrac{\xi}{2}},
\end{aligned}
\right.\label{MHCPPPDF}
\end{align}
where $r_\mr{m}>0$ denotes the distance from the PPP typical UE to MHCPP gNB; $\varrho\approx0.3686(\lambda_\mr{m}\pi\xi^2)^2+0.0985\lambda_\mr{m}\pi\xi^2+2$ based on the simulation in \cite{8060551}.
\begin{IEEEproof}
A detailed proof can be found in \cite{8060551}.
\end{IEEEproof}

Given the statistics of the contact distance, we now derive the probabilities of the typical UE associated with an LoS IAB-node, an NLoS IAB-node, and a gNB in the following propositions.

\textit{Proposition 2: The probability of a typical UE associated with an LoS/NLoS IAB-node at $\mathbf{x}_\mr{s}^\mr{L}$/$\mathbf{x}_\mr{s}^\mr{N}$, denoted as $\mathcal{A}_{\mr{s},\mr{L}}$/$\mathcal{A}_{\mr{s},\mr{N}}$, is expressed as:
\begin{align}
    \mathcal{A}_{\mr{s},i}&=\int_0^{\left(\tfrac{\xi}{2\Delta_{i,1}}\right)^{\tfrac{\alpha_\mr{m}}{\alpha_{\mr{s},i}}}}\left(1-F_{R_{\mathbf{x}_\mr{m}\mathbf{0}},1}\left(r_{\mr{s},i}^{\tfrac{\alpha_{\mr{s},i}}{\alpha_\mr{m}}}\Delta_{i,1}\right)\right)\notag\\&\times\left(1-F_{R_{\mathbf{x}_\mr{s}^j\mathbf{0}}}\left(r_{\mr{s},i}^{\tfrac{\alpha_{\mr{s},i}}{\alpha_{\mr{s},j}}}\Delta_{i,2}\right)\right)f_{R_{\mathbf{x}_\mr{s}^i\mathbf{0}}}(r_{\mr{s},i})\mr{d}r_{\mr{s},i}\notag\\&+\int^\infty_{\left(\tfrac{\xi}{2\Delta_{i,1}}\right)^{\tfrac{\alpha_\mr{m}}{{\alpha_{\mr{s},i}}}}}\left(1-F_{R_{\mathbf{x}_\mr{m}\mathbf{0}},2}\left(r_{\mr{s},i}^{\tfrac{\alpha_{\mr{s},i}}{\alpha_\mr{m}}}\Delta_{i,1}\right)\right)\notag\\&\times\left(1-F_{R_{\mathbf{x}_\mr{s}^j\mathbf{0}}}\left(r_{\mr{s},i}^{\tfrac{\alpha_{\mr{s},i}}{\alpha_{\mr{s},j}}}\Delta_{i,2}\right)\right)f_{R_{\mathbf{x}_\mr{s}^i\mathbf{0}}}(r_{\mr{s},i})\mr{d}r_{\mr{s},i},\label{asl}
\end{align}
where $i=\mr{L}$, $j=\mr{N}$; $i=\mr{N}$, $j=\mr{L}$. $\Delta_{i,1}=\left(\tfrac{P_\mr{m}N_\mr{T}^\mr{m}N^2_\mr{s}T_\mr{m}}{P_\mr{s}N_\mr{T}^\mr{s}N_\mr{m}^2T_\mr{s}}\right)^{\tfrac{1}{\alpha_\mr{m}}}e^{{\tfrac{\left(0.1\zeta_\mr{m}\mr{ln}10\right)^2-\left(0.1\zeta_{\mr{s},i}\mr{ln}10\right)^2}{2\alpha_\mr{m}}}}$ and $\Delta_{i,2}=$ $e^{{\tfrac{\left(0.1\zeta_{\mr{s},j}\mr{ln}10\right)^2-\left(0.1\zeta_{\mr{s},i}\mr{ln}10\right)^2}{2\alpha_{\mr{s},j}}}}$. The integration bound is given by solving $r_{\mr{s},i}^{\tfrac{\alpha_{\mr{s},i}}{\alpha_\mr{m}}}\Delta_{i,1}=\tfrac{\xi}{2}$.}

\begin{IEEEproof}
The proof can be found in Appendix-\ref{app2}.
\end{IEEEproof}

\textit{Proposition 3: The probability of a typical UE associated with a gNB at $\mathbf{x}_\mr{m}$, represented as $\mathcal{A}_\mr{m}$, is expressed as: 
\begin{align}
    \mathcal{A}_\mr{m}=&\int_0^{\tfrac{\xi}{2}}\left(1-F_{R_{\mathbf{x}_\mr{s}^\mr{L}\mathbf{0}}}\left(r_{\mr{m}}^{\tfrac{\alpha_\mr{m}}{\alpha_{\mr{s},\mr{L}}}}\Delta_{\mr{L}}\right)\right)\notag\\&\times\left(1-F_{R_{\mathbf{x}_\mr{s}^\mr{N}\mathbf{0}}}\left(r_{\mr{m}}^{\tfrac{\alpha_\mr{m}}{\alpha_{\mr{s},\mr{N}}}}\Delta_{\mr{N}}\right)\right)f_{R_{\mathbf{x}_\mr{m}\mathbf{0}},1}(r_\mr{m})\mr{d}r_\mr{m}\notag\\&+\int^\infty_{\tfrac{\xi}{2}}\left(1-F_{R_{\mathbf{x}_\mr{s}^\mr{L}\mathbf{0}}}\left(r_{\mr{m}}^{\tfrac{\alpha_\mr{m}}{\alpha_{\mr{s},\mr{L}}}}\Delta_{\mr{L}}\right)\right)\notag\\&\times\left(1-F_{R_{\mathbf{x}_\mr{s}^\mr{N}\mathbf{0}}}\left(r_{\mr{m}}^{\tfrac{\alpha_\mr{m}}{\alpha_{\mr{s},\mr{N}}}}\Delta_{\mr{N}}\right)\right)f_{R_{\mathbf{x}_\mr{m}\mathbf{0}},2}(r_\mr{m})\mr{d}r_\mr{m},\label{am}
\end{align}
where $\Delta_{i}=\left(\tfrac{P_\mr{s}N_\mr{T}^\mr{s}N_\mr{m}^2T_\mr{s}}{P_\mr{m}N_\mr{T}^\mr{m}N_\mr{s}^2T_\mr{m}}\right)^{\tfrac{1}{\alpha_{\mr{s},i}}}e^{\tfrac{\left(0.1\zeta_{s,i}\mr{ln}10\right)^2-\left(0.1\zeta_m\mr{ln}10\right)^2}{2\alpha_{\mr{s},i}}}$ for $i\in\{\mr{L},\mr{N}\}$.}

\begin{IEEEproof}
The proof follows similar steps to Proposition 2 given the condition that the nearest gNB provides higher biased-received-desired-signal power than the nearest LoS and NLoS IAB-nodes.
\end{IEEEproof}

Having obtained the association probabilities, we can obtain the PDFs of the serving distance (i.e., the distance of the typical UE to its serving gNB or IAB-node) in the following proposition.

{\textit{Proposition 4: Conditioned on the typical UE is served by its nearest gNB at $\mathbf{x}_\mr{m}$, and the corresponding PDF of the serving distance is written as
\begin{align}
\hat{f}_{R_{\mathbf{x}_\mr{m}\mathbf{0}}}(r_\mr{m})&=\tfrac{1}{\mathcal{A}_\mr{m}}\left(1-F_{R_{\mathbf{x}_\mr{s}^\mr{L}\mathbf{0}}}\left(r_{\mr{m}}^{\tfrac{\alpha_\mr{m}}{\alpha_{\mr{s},\mr{L}}}}\Delta_{\mr{L}}\right)\right)\notag\\&\times\left(1-F_{R_{\mathbf{x}_\mr{s}^\mr{N}\mathbf{0}}}\left(r_{\mr{m}}^{\tfrac{\alpha_\mr{m}}{\alpha_{\mr{s},\mr{N}}}}\Delta_{\mr{N}}\right)\right) f_{R_{\mathbf{x}_\mr{m}\mathbf{0}}}(r_\mr{m}),
\label{sss}
\end{align}
where $f_{R_{\mathbf{x}_\mr{m}\mathbf{0}}}(r_\mr{m})$ is piece-wise at point $r_\mr{m}=\tfrac{\xi}{2}$ as stated in Lemma 2.}

\textit{Similarly, conditioned on the typical UE being served by its nearest LoS (or NLoS) IAB-node at $\mathbf{x}_\mr{s}^\mr{L}$ (or $\mathbf{x}_\mr{s}^\mr{N}$), the corresponding PDF of the serving distance is given as
\begin{align}
\hat{f}_{R_{\mathbf{x}_\mr{s}^i\mathbf{0}}}(r_{\mr{s},i})&=\tfrac{1}{\mathcal{A}_{\mr{s},i}}\left(1-F_{R_{\mathbf{x}_\mr{m}\mathbf{0}}}\left(r_{\mr{s},i}^{\tfrac{\alpha_{\mr{s},i}}{\alpha_\mr{m}}}\Delta_{i,1}\right)\right)\notag\\&\times\left(1-F_{R_{\mathbf{x}_\mr{s}^j\mathbf{0}}}\left(r_{\mr{s},i}^{\tfrac{\alpha_{\mr{s},i}}{\alpha_{\mr{s},j}}}\Delta_{i,2}\right)\right)f_{R_{\mathbf{x}_\mr{s}^i\mathbf{0}}}(r_{\mr{s},i}).
\label{ss}
\end{align}
When $i=\mr{L}$, we have $j=\mr{N}$; and when $i=\mr{N}$, we have $j=\mr{L}$. According to Lemma2, $F_{R_{\mathbf{x}_\mr{m}\mathbf{0}}}\left(r_{\mr{s},i}^{\tfrac{\alpha_{\mr{s},i}}{\alpha_\mr{m}}}\Delta_{i,1}\right)$ is piece-wise at point $r_{\mr{s},i}={\left(\tfrac{\xi}{2\Delta_{i,1}}\right)^{\tfrac{\alpha_\mr{m}}{\alpha_{\mr{s},i}}}}$ by solving $r_{\mr{s},i}^{\tfrac{\alpha_{\mr{s},i}}{\alpha_\mr{m}}}\Delta_{i,1}=\tfrac{\xi}{2}$.}}
\begin{IEEEproof}
A simple proof can be found in \cite[Lemma 4]{8882288}.
\end{IEEEproof}

In the remaining subsections, we derive some performance metrics, such as $\mr{SINR}$ coverage, capacity with outage, and ergodic capacity for single-hop backhaul mmWave-IBFD/HD-IAB networks. We set the available bandwidth for IBFD communications as $W$. Assume that the HD transmission is in a frequency-division-duplexing (FDD) manner, whose available bandwidth is equally divided into two orthogonal portions (i.e., $W/2$): one for the gNB tier and the other for the IAB-node tier.

\subsection{$\mr{SINR}$ Coverage}
We first give the expression on the $\mr{SINR}$ coverage in the following theorem.

\textit{Theorem 1: The $\mr{SINR}$ coverage of the single-hop backhaul mmWave-IBFD-IAB network is defined as
\begin{align}
    \mathcal{P}_\mr{FD}(\tau)&=\mathcal{A}_\mr{m}\mathbb{P}\left(\mr{SINR}_{\mathbf{0}\mathbf{x}_\mr{m}}>\tau\right)\notag\\&+\sum_{i\in\{\mr{L},\mr{N}\}}\mathcal{A}_{\mr{s},i}\mathbb{P}\left(\mr{SINR}_{\mathbf{0}\mathbf{x}_\mr{s}^i}>\tau,\mr{SINR}_{\mathbf{x}_\mr{s}^i\mathbf{x}_\mr{m}}>\tau\right),\label{sc}
\end{align}
where $\tau$ denotes the target $\mr{SINR}$. $\mr{SINR}_{\mathbf{x}_\mr{s}^i\mathbf{x}_\mr{m}}$ denotes the backhaul link $\mr{SINR}$ of the IAB-node at $\mathbf{x}_\mr{s}^i$ for serving a typical UE. However, solving $\mathbb{P}\left(\mr{SINR}_{\mathbf{0}\mathbf{x}_\mr{s}^i}>\tau,\mr{SINR}_{\mathbf{x}_\mr{s}^i\mathbf{x}_\mr{m}}>\tau\right)$ is intractable. Fortunately, since IAB-nodes and UEs are drawn from homogeneous PPP, according to Slivnyak’s theorem, we have \cite{8882288,7110547}
\begin{align}
    \mathbb{P}\left(\mr{SINR}_{\mathbf{0}\mathbf{x}_\mr{s}^i}>\tau,\mr{SINR}_{\mathbf{x}_\mr{s}^i\mathbf{x}_\mr{m}}>\tau\right)&=\mathbb{P}\left(\mr{SINR}_{\mathbf{0}\mathbf{x}_\mr{s}^i}>\tau\right)\notag\\&\times\mathbb{P}\left(\mr{SINR}_{\hat{\mathbf{0}}\mathbf{x}_\mr{m}}>\tau\right).\label{simp}
\end{align}
The solution of this theorem can be found in Appendix-\ref{appB}.}

The $\mr{SINR}$ coverage of the single-hop backhaul mmWave-HD-IAB network $\mathcal{P}_\mr{HD}(\tau)$ can be derived similarly to Theorem 1; however, owing to the FDD, the interference from the other tier and the RSI in the $\mr{SINR}$ expressions should be removed. Moreover, the Gaussian noise power of HD is reduced by half compared to that of IBFD because the bandwidth of IBFD is twice as large as that of HD.

\subsection{Capacity with Outage}
In the following theorem, we derive the expression on the capacity with outage.

\textit{Theorem 2: The capacity with outage, that is, the average rate correctly received over many transmission bursts, of the single-hop backhaul mmWave-IBFD-IAB network is cast as
\begin{align}
    \mathcal{C}_\mr{FD}(\tau_\mr{min})=W\log_2\left(1+\tau_\mr{min}\right)\mathcal{P}_\mr{FD}(\tau_\mr{min}),\label{cwof}
\end{align}
where $\tau_\mr{min}$ is the minimum received $\mr{SINR}$ fixed at the transmitter because the transmitter does not know the instantaneous received $\mr{SINR}$; thus, the transmission rate should be fixed independently.}

Similarly, the capacity with outage of the single-hop backhaul mmWave-HD-IAB network $\mathcal{C}_\mr{HD}(\tau_\mr{min})$ is given by replacing $W$ with $W/2$ and $\mathcal{P}_\mr{FD}(\tau_\mr{min})$ with $\mathcal{P}_\mr{HD}(\tau_\mr{min})$ in Theorem 2.

\subsection{Ergodic Capacity}
The expression on the ergodic capacity is presented in the following theorem.

\textit{Theorem 3: The ergodic capacity for single-hop backhaul mmWave-IBFD-IAB networks, defined as the average of the instantaneous capacity, is written as
\begin{align}
    \mathcal{E}_\mr{FD}&=\mathcal{A}_\mr{m}{\bar{\mathcal{R}}_\mr{m}}+\mathcal{A}_{\mr{s},\mr{L}}\min\left({\bar{\mathcal{R}}_{\mr{s},\mr{L}}},{\bar{\mathcal{R}}_\mr{b}}\right)\notag\\&+\mathcal{A}_{\mr{s},\mr{N}}\min\left({\bar{\mathcal{R}}_{\mr{s},\mr{N}}},{\bar{\mathcal{R}}_\mr{b}}\right),
\end{align}
where $\bar{\mathcal{R}}_\mr{m}=\mathbb{E}\{W\log_2(1+\mr{SINR}_{\mathbf{0}\mathbf{x}_\mr{m}})\}$, $\bar{\mathcal{R}}_{\mr{s},\mr{L}}=\mathbb{E}\{W\log_2(1+\mr{SINR}_{\mathbf{0}\mathbf{x}_\mr{s}^{\mr{L}}})\}$, $\bar{\mathcal{R}}_{\mr{s},\mr{N}}=\mathbb{E}\{W\log_2(1+\mr{SINR}_{\mathbf{0}\mathbf{x}_\mr{s}^{\mr{N}}})\}$ and $\bar{\mathcal{R}}_\mr{b}=\mathbb{E}\{W\log_2(1+\mr{SINR}_{\hat{\mathbf{0}}\mathbf{x}_\mr{m}})\}$ represent ergodic capacities of the gNB associated access link, LoS IAB-node associated access link, NLoS IAB-node associated access link, and backhaul link, respectively, which are derived in Appendix-\ref{D}.}

Likewise, the ergodic capacity of the corresponding HD network $\mathcal{E}_\mr{HD}$ is given by scaling $W$ in Theorem 3 with $1/2$ and removing the RSI and interference from the other tier in $\mr{SINR}$ expressions.

Note that we choose the minimum ergodic capacity between the IAB-node associated access link and the backhaul link as the final ergodic capacity at the UE associated with the IAB-node. This is because, in the IAB system, the maximum ergodic capacity that an IAB-node can provide is limited by its corresponding backhaul link \cite{7110547,8493520}.

\begin{table*}[!t]
\renewcommand{\arraystretch}{0.8}
%\extrarowheight %as needed to properly center the text within the cells
\caption{System parameters and default values}
\label{notation}
\centering
\begin{threeparttable}
\begin{tabular}{|c|c|c|}
\hline
\textbf{Notation} & \textbf{Physical Meaning} & \textbf{Values}\\
\hline
$K$ & Number of subcarriers & 512 {\cite{ieee}}\\
\hline
$W$ & Bandwidth & 800 MHz\\
\hline
$f_\mr{c}$ & Carrier frequency & 38 GHz\\
\hline
$M_\mr{m},M_{\mr{s},\mr{L}},M_{\mr{s},\mr{N}}$ & Nakagami factor & 4, 3, 2\\
\hline
$\sigma_\mr{n}^2$ & Gaussian noise power& 
\tabincell{l}{$-174\;\mr{dBm}+10\log_{10}W+10$ dB (IBFD)\\$-174\;\mr{dBm}+10\log_{10}\tfrac{W}{2}+10$ dB (HD)}\\
\hline
$\epsilon$ & Blockage density constant & 0.0071{\cite{6932503}}\\
\hline
$R_0$ & Radius of analysis region & 1000m\\
\hline
$\xi$ & Hard-core distance & 100m\\
\hline
${\lambda}_\mr{m},\lambda_\mr{s},{\lambda}_\mr{u}$ & gNB, IAB-node, and UE density & $1\times10^{-5}/\mr{m}^2$, $4\times10^{-5}/\mr{m}^2$, $2\times10^{-4}/\mr{m}^2$\\
\hline
$\eta$ & RSI controlling factor & $80$ dB {\cite{Zhan2012}}\\ 
\hline
$\eta_\mr{dig}$ & The amount of digital SIC & $25$ dB {\cite{Zhan2012}}\\ 
\hline
$T$ & Gamma distribution samples & $5$ {\cite{7857035}}\\ 
\hline
${P_\mr{m}},{P_\mr{s}}$ & Transmit power & $40$ dBm, $33$ dBm\\
\hline
${T_\mr{s}}/{T_\mr{m}}$ & Bias factor ratio& 0 dB\\
\hline
${N_\mr{m}},{N_\mr{s}}$ & Number of subarrays & $8$, $4$\\
\hline
${N_\mr{T}^\mr{m}},{N_\mr{T}^\mr{s}}$ & Number of transmit antennas & $32\times32$, $16\times16$\\
\hline
${N_\mr{R}^\mr{s}},{N_\mr{R}^\mr{u}}$ & Number of receive antennas & $16\times16$, $8\times8$\\
\hline
$\alpha_\mr{m},\alpha_{\mr{s},\mr{L}},\alpha_{\mr{s},\mr{N}}$ & Path loss exponent & 1.9, 2, 3.3 {\cite{6894456}}\\ 
\hline
$\zeta_\mr{m},\zeta_{\mr{s},\mr{L}},\zeta_{\mr{s},\mr{N}}$ & Shadowing effect standard deviation & 3.7, 4.3, 10.7 {\cite{6894456}}\\
\hline
$q$ & ADC quantization bits & 8 bits\\
\hline
$\hat{g}_\mr{T,m},\hat{g}_\mr{T,s},{g}_\mr{T,m},{g}_\mr{T,s}$ & Transmitter sidelobe gain & -8.9 dB, -5.9 dB, -17.8 dB, -11.7 dB {\cite{antenna,7434599}}\\
\hline
$\hat{g}_\mr{R,s},\hat{g}_\mr{R,u},{g}_\mr{R,s},{g}_\mr{R,u}$ & Receiver sidelobe gain & -5.9 dB, -2.9 dB, -11.7 dB, -5.7 dB {\cite{antenna,7434599}}\\
\hline
\end{tabular}
\end{threeparttable}
\end{table*}
\section{Numerical Results}\label{simulation}
In this section, we verify the theoretical results derived in this study and evaluate the performance of the single-hop backhaul mmWave-IBFD-IAB networks. Furthermore, to better understand the advantages and disadvantages of IBFD transmission, comparisons are made between IBFD and HD schemes. Monte Carlo simulations with $10^6$ iterations are performed to verify the accuracy of the derived theoretical results. Unless otherwise specified, the values of the simulation parameters are presented in Table~\ref{notation}.

\subsection{Association Probabilities and $\mr{SINR}$ Coverage}
\begin{figure}[t!]
\centering
\subfigure[]{
\includegraphics[width=\columnwidth]{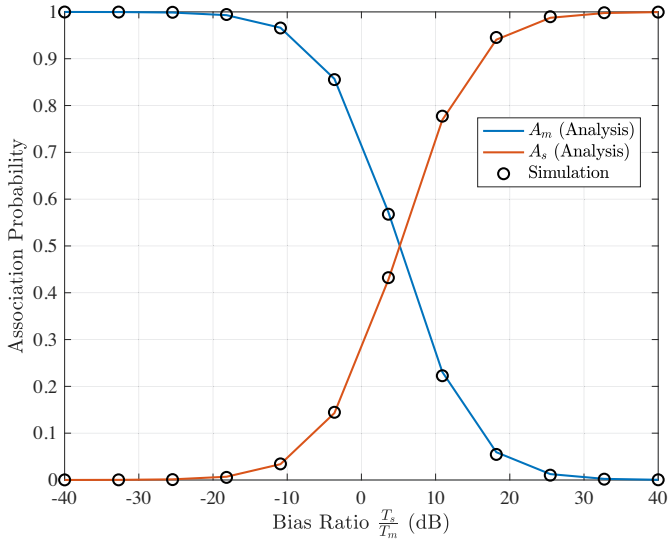}\label{ver1}}
\subfigure[]{\includegraphics[width=\columnwidth]{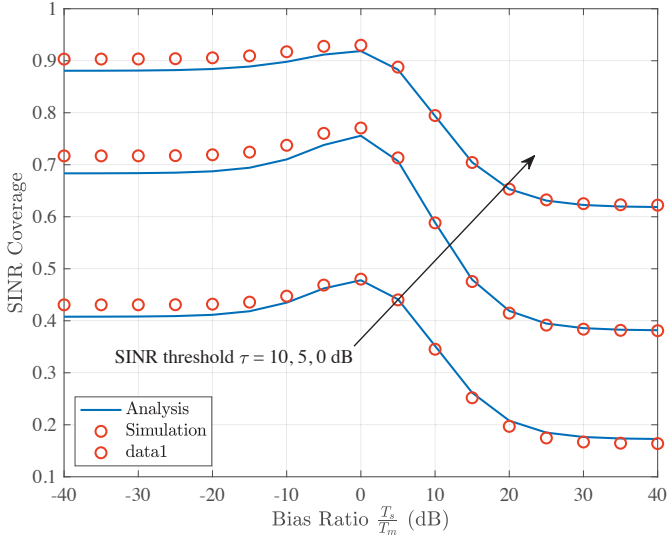}\label{ver2}}
\caption{Verification of theoretical results for (a) association probabilities and (b) $\mr{SINR}$ coverage v.s. bias ratio $\tfrac{T_\mr{s}}{T_\mr{m}}$ at $\mr{SINR}$ threshold $\tau=0,5,10$ dB.}
\label{ver}
\end{figure}

Both figures in Fig.~\ref{ver} show a close match between the simulation and analytical results, which indicates good accuracy given by our approximated analytical results derived in this study. In Fig.~\ref{ver1}, we evaluate how the association probabilities can be affected by changing the bias ratio of the IAB-node to the gNB, that is, $\tfrac{T_\mr{s}}{T_\mr{m}}$. We can find that the association probability of a typical UE associates with an IAB-node, that is, $\mathcal{A}_\mr{s}$, increases as the ratio increases, and that of the typical UE associates with a gNB, that is, $\mathcal{A}_\mr{m}$, shows an opposite trend because as $\tfrac{T_\mr{s}}{T_\mr{m}}$ increases, more traffic is offloaded from the gNB to the IAB-node. 

In Fig.~\ref{ver2}, we study the impact of $\tfrac{T_\mr{s}}{T_\mr{m}}$ on $\mr{SINR}$ coverage. The $\mr{SINR}$ coverage at a fixed threshold first increases as $\tfrac{T_\mr{s}}{T_\mr{m}}$ increases; after reaching the maximum at $\tfrac{T_\mr{s}}{T_\mr{m}}=0$ dB, the $\mr{SINR}$ coverage decreases as the ratio increases. The reasons for this decrease are as follows. As $\tfrac{T_\mr{s}}{T_\mr{m}}$ increases, $\mathcal{A}_\mr{s}$ becomes dominant, as shown in Fig.~\ref{ver1}. Thus, according to Theorem 1, the $\mr{SINR}$ coverage of the network can be dominated by that of the IAB-node tier at a high bias ratio. Even if the gNB is tower-mounted, it has a direct fiber connection but a limited quantity owing to the high deployment cost of fiber-connected BSs. Biasing towards wireless IAB-nodes has the potential benefit of reducing deployment costs. However, because i) the transmit power and propagation properties of the IAB-node are worse than those of the gNB; ii) the performance of the IAB-node associated access link is restricted by the backhaul link, and iii) the backhaul link suffers from RSI, the network coverage becomes distorted. Moreover, it is interesting to note that the rate of increase in coverage is much lower than the decrement rate.

\subsection{Capacity with Outage}
\begin{figure}[t!]
\centering
\subfigure[]{
\includegraphics[width=\columnwidth]{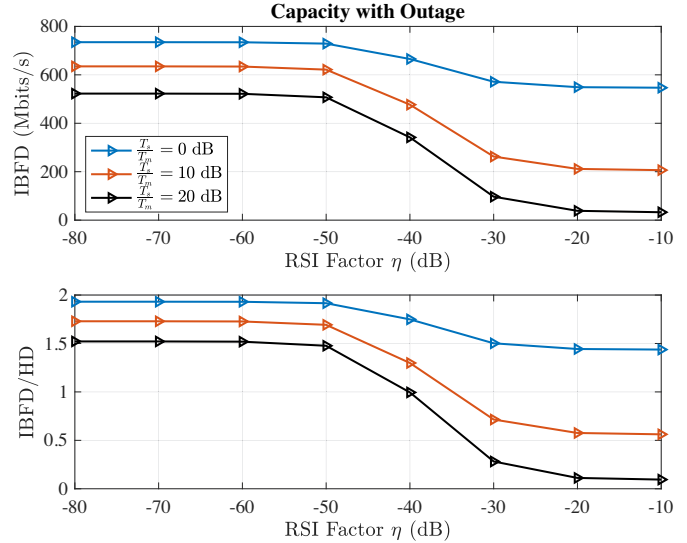}\label{rsi}}
\subfigure[]{
\includegraphics[width=\columnwidth]{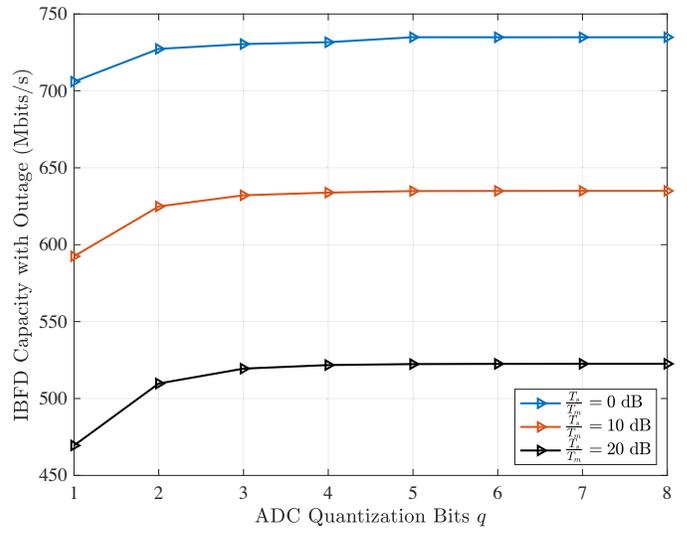}\label{qua}}
\caption{Single-hop backhaul mmWave-IBFD-IAB networks capacity with outage at a minimum received $\mr{SINR}=0$ dB (i.e., $\tau_\mr{min}=0$ dB) in terms of $\tfrac{T_\mr{s}}{T_\mr{m}}=0,10,20$ dB (a) v.s. different values of RSI factors; (b) v.s. different values of ADC quantization bits.}
\end{figure}

We assume that the transmitter fixes the minimum received $\mr{SINR}$ at $\tau_\mr{min}=0$ dB. In Fig.~\ref{rsi}, we compare the capacity with outage of single-hop backhaul mmWave-IBFD- and HD-IAB networks in terms of different bias ratios $\tfrac{T_\mr{s}}{T_\mr{m}}$ {with various values of the RSI controlling factor $\eta$, which denotes the level of SIC (higher values denote poorer SIC)}. In general, as $\tfrac{T_\mr{s}}{T_\mr{m}}$ increases, the capacity with coverage decreases, which can be explained by Theorem 2 and the analysis in previous subsection. When $\eta>-50$ dB, the effect of the RSI on the capacity with outage becomes obvious. However, when $\eta>-20$ dB, the capacity with outage for different values of $\tfrac{T_\mr{s}}{T_\mr{m}}$ tends to be stable. This is because a high RSI makes the $\mr{SINR}$ coverage of the IAB-node tier approach 0; hence, the $\mr{SINR}$ coverage of the network is nearly dominated by that of the gNB tier. {In addition, a trade-off between using IBFD and HD can be observed in the figure because the capacity with outage ratio of IBFD to HD decreased as $\eta$ and/or $\tfrac{T_\mr{s}}{T_\mr{m}}$ increased. In some circumstances, HD can outperform the IBFD. For example, for $\tfrac{T_\mr{s}}{T_\mr{m}}=20$ dB, HD provides a higher capacity with outage than IBFD when $\eta>-40$ dB. Interestingly, with efficient SIC, for $\tfrac{T_\mr{s}}{T_\mr{m}}=0$ dB, using IBFD transmission can nearly double the capacity with outage compared to using HD; even if a high RSI is reserved, the capacity with outage of IBFD still outperforms that of HD.} This is an attractive result because one can benefit from IBFD with high RSI power by tuning $\tfrac{T_\mr{s}}{T_\mr{m}}$ rather than exploring difficult and expensive SIC techniques to achieve higher levels of SIC. In addition, the capacity with outage ratio of IBFD to HD decreased as $\tfrac{T_\mr{s}}{T_\mr{m}}$ increased.

In Fig.~\ref{qua}, we also study how the capacity with outage of the single-hop backhaul mmWave-IBFD-IAB network is affected when using an ADC with different resolutions at various bias ratios at $\tau_\mr{min}=0$ dB. Evidently, the capacity with outage can be improved by increasing the ADC quantization bits and reducing the bias ratio. Meanwhile, the capacity with outage saturates when the ADC resolution is higher than 5 bits, indicating that the ADC quantization noise effect becomes negligible compared to other interference and noise in our analyzed network. This result emphasizes the feasibility of using a low-resolution ADC in a network. Low-resolution ADCs are preferred in mmWave communications because their low power consumption can compensate for the high power consumption imposed by numerous RF chains \cite{8337813}.

The simulation results for verifying the analytical results of capacity with outage are omitted because they depend on the $\mr{SINR}$ coverage, which shows a close match between the simulation and analytical results in Fig.~\ref{ver2}.

\subsection{Ergodic Capacity}
\begin{figure}[t!]
\centering
\subfigure[]{
\includegraphics[width=\columnwidth]{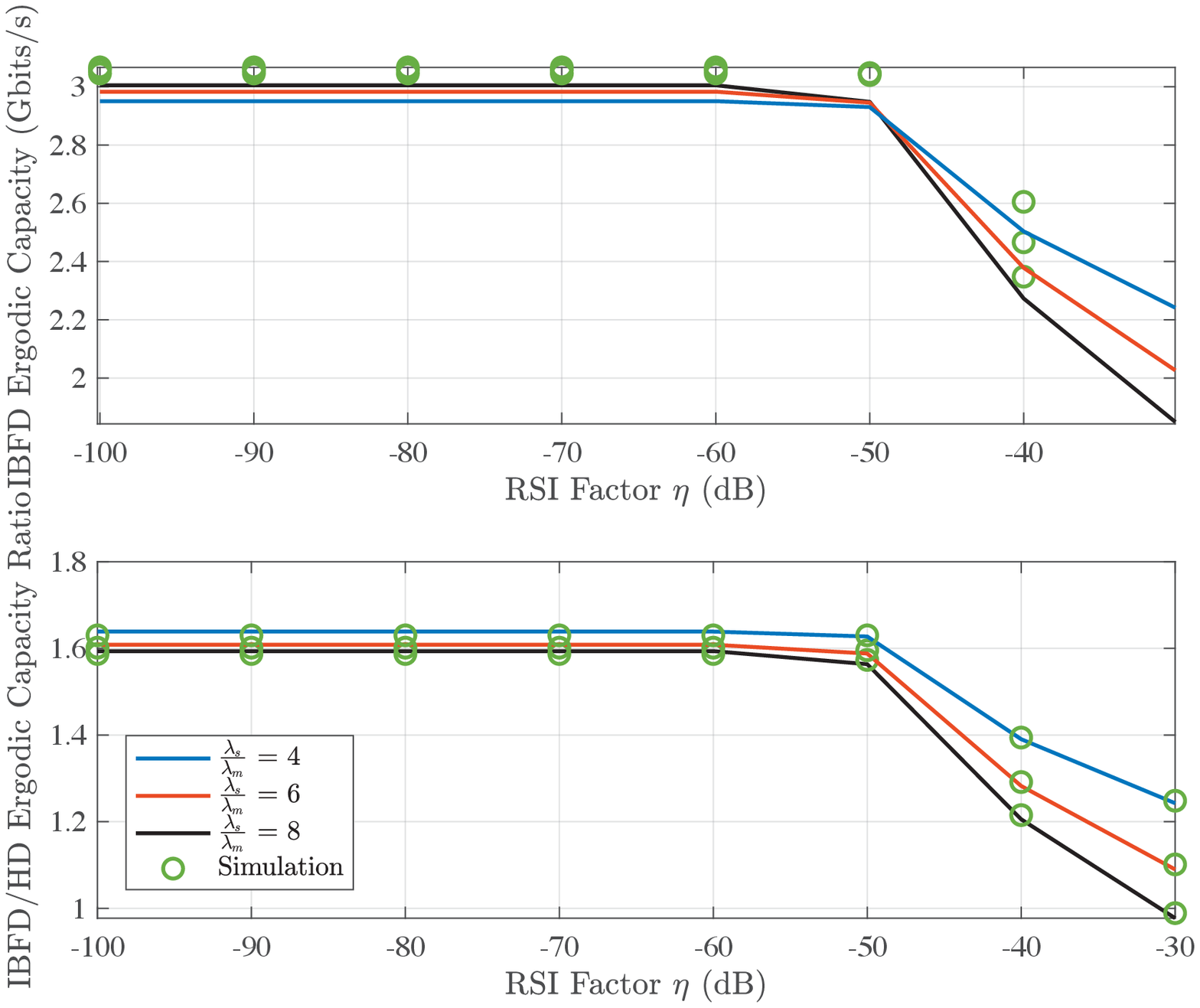}\label{latency1}}
\subfigure[]{
\includegraphics[width=\columnwidth]{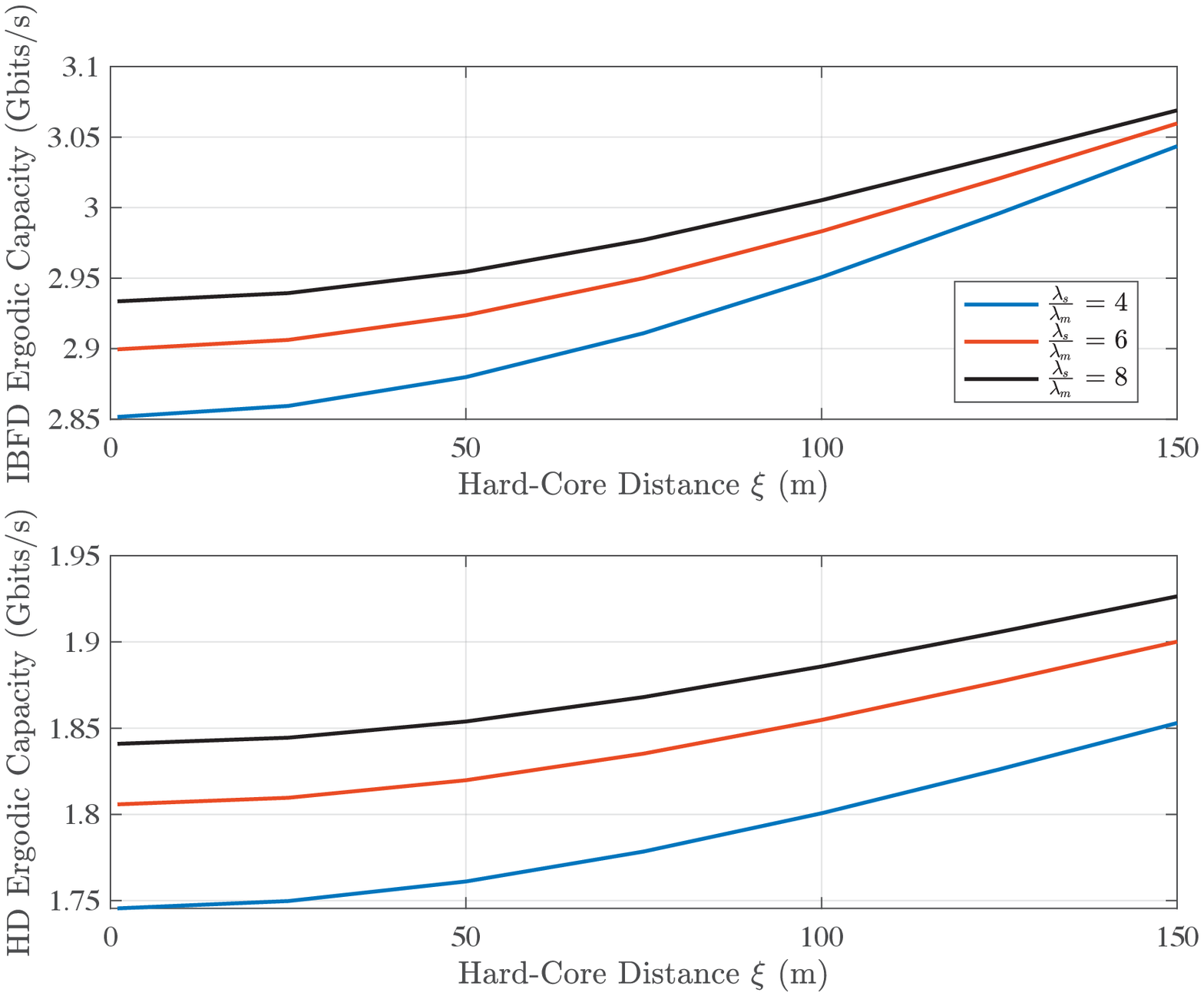}\label{latency2}}
\caption{Single-hop backhaul mmWave-IBFD-IAB networks ergodic capacity in terms of $\tfrac{\lambda_\mr{s}}{\lambda_\mr{m}}=4,6,8$ (${\lambda_\mr{m}}=1\times10^{-5}/\mr{m}^2$) (a) v.s. different RSI factor; (b) v.s. different hard-core distance.}
\label{la}
\end{figure}

Fig.~\ref{la} compares the network ergodic capacity with varying RSI factor $\eta$ and hard-core distance $\xi$ between single-hop backhaul FR2-IBFD- and HD-IAB networks in terms of different density ratios of the IAB-node to gNB $\tfrac{\lambda_\mr{s}}{\lambda_\mr{m}}$. Fig.~\ref{latency1} shows how the RSI affects the network ergodic capacity. Note that when $\tfrac{\lambda_\mr{s}}{\lambda_\mr{m}}$ increases, the effective cell radius of the IAB-node tier decreases, resulting in a higher ergodic capacity of the corresponding link. In contrast, the ergodic capacity of the gNB tier is reduced because the UEs are more likely to be served by the IAB-nodes as $\tfrac{\lambda_\mr{s}}{\lambda_\mr{m}}$ increases. When $\eta\leq-50$ dB, the ergodic capacity increment of the IAB-node tier is higher than the decrement of the gNB tier as the ratio increases. Thus, the overall ergodic capacity increases as the ratio increases. However, for $\eta>-50$ dB, the growth in the ergodic capacity of the IAB-node tier is less than that of the gNB tier as the ratio increases; hence, the overall ergodic capacity decreases as the ratio increases. This result indicates that with successful SIC, the network ergodic capacity can be improved by increasing $\tfrac{\lambda_\mr{s}}{\lambda_\mr{m}}$. Compared with HD transmission, if $\eta\leq-50$ dB, the ergodic capacity provided by the IBFD scheme is around 1.6 times higher than that provided by HD transmission, regardless of the density ratio. {However, in the high RSI factor $\eta$ region, the higher the density ratio, the lower the ergodic capacity provided by the IBFD, and thus, the lower the ergodic capacity ratio of IBFD to HD $\tfrac{\mathcal{E}_\mr{FD}}{\mathcal{E}_\mr{HD}}$ is. Moreover, a similar trade-off between using HD and IBFD, as shown in Fig.~\ref{rsi}, can be concluded here, i.e., the ergodic capacity of HD can outperform IBFD for a high $\tfrac{\lambda_\mr{s}}{\lambda_\mr{m}}$ and/or high $\eta$.} Since the close match between the simulation and analytical results is shown in the figure, which indicates good accuracy given by our analytical results derived for the network ergodic capacity.

In Fig.~\ref{latency2}, by fixing the density of the gNB, the interference given by the gNBs is suppressed by increasing the hard-core distance $\xi$, and as $\xi$ increases, the network ergodic capacity increases, which is the case for both the IBFD and HD schemes. In particular, as $\xi$ increases, the IBFD ergodic capacity gap between different node density ratios is reduced. Interestingly, when $\xi$ is equal to 0, MHCPP converges to PPP, which yields the lowest ergodic capacity. This result highlights the necessity of imposing a suitable distance between the gNBs. Moreover, with successful SIC, IBFD provides a higher ergodic capacity than HD, regardless of $\xi$ is. The simulation results are omitted here because the accuracy of our analytical results on the network ergodic capacity is verified in Fig.~\ref{latency1}.

\subsection{Effect of Assumption 1}
\begin{figure}[t!]
\centering
\includegraphics[width=\columnwidth]{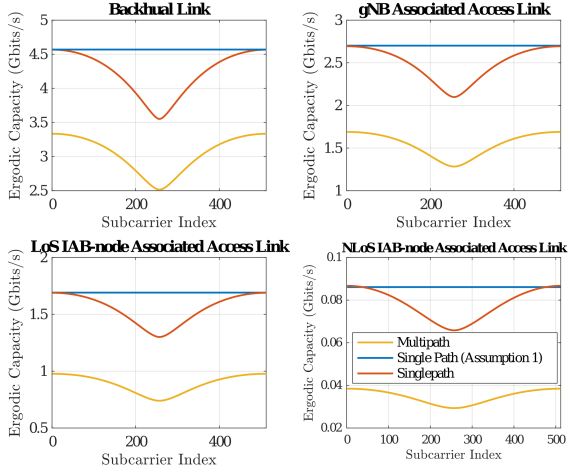}
\caption{Ergodic capacity comparison for different links per subcarrier in terms of single path and multi-path scenario.}
\label{effect}
\end{figure}

In Fig.~\ref{effect}, we simulate the ergodic capacity for different links per subcarrier to analyze the effect of Assumption 1 by simulations and provide the corresponding ergodic capacity for the multi-path scenario with three paths for both the LoS and NLoS. For the multi-path scenario, we relax Assumption 1, and the power of the desired signal is calculated based on Lemma 1 in \cite{7434599}. This lemma states that for a large number of antennas, when the number of paths is much smaller than the number of antennas, the power of the desired signal converges to the power given by the strongest path, where the optimal RF beamformers are given by the AoA and AoD of the strongest path. With Assumption 1, the single path channel is assumed to exhibit flat fading, that is, the channel gain is the same for all subcarriers. Therefore, we obtain a constant ergodic capacity across all subcarriers. By relaxing Assumption 1, $\gamma[k]$ in \eqref{channel} varies for each subcarrier, which gives us the convex curves in Fig.~\ref{effect}. It can be seen that Assumption 1 is a strong assumption, which gives us an upper band on ergodic capacity. Without Assumption 1, the performance analyzed in the previous subsections is degraded. Moreover, by adding a multi-path to the picture, we can observe further degradation of ergodic capacity. This is because multi-path propagation introduces additional interference, which reduces the $\mr{SINR}$ compared with that in the single path scenario. However, even if Assumption 1 is a strong assumption, it provides tractability in stochastic geometry analysis for OFDM modulation, because the statistics of raised cosine pulse shaping filters are still unknown in the literature and are difficult to derive owing to the complex impulse response of the filter \cite{glover2010digital,4534833}.

\section{Conclusion}
In this study, the performance of multi-cell wideband single-hop backhaul mmWave-IBFD-IAB networks using stochastic geometry, where the gNBs are deployed in MHCPP, was investigated. For the Nakagami-$M$ small fading and lognormal shadowing effect, we leveraged the approximated composite GL distribution for network performance analysis. Moreover, for simplicity, the PGFL of PPP was adopted to approximate the mean interference given by the gNBs. Consequently, our closed-form performance metrics showed a close match with the simulations. Furthermore, comparisons were made with the performance of the single-hop backhaul mmWave-HD-IAB networks. The numerical results demonstrated that tuning the bias ratio of the IAB-node to gNB, the $\mr{SINR}$ coverage at a fixed $\mr{SINR}$ threshold shows a convex trend. Interestingly, we found that by selecting a suitable bias ratio, the IBFD scheme can always perform better than HD, even with a high RSI power, relaxing the requirement for high-quality SIC techniques. Moreover, at $\mr{SINR}=0$ dB, the effect of ADC quantization noise can be neglected at a lower resolution. As for the network ergodic capacity, fixing the density of gNB, the ergodic capacity was enhanced by deploying gNBs with MHCPP compared to deploying gNBs with PPP, emphasizing the advantage of utilizing MHCPP. Moreover, the larger the hard-core distance, the higher the ergodic capacity. Additionally, with a small RSI power, the ergodic capacity can be improved by increasing the ratio of the density of the IAB-node to gNB.

Further work can be focused on determing the transformation properties of the MHCPP and deriving the system performance with the impact of hardware impairments and channel estimation error. Moreover, the design and analysis of a multi-hop backhaul scenario are worth studying.

\appendices
\section{Composite Gamma-Lognormal Distribution}\label{GL}
Given that $X\sim \mr{Gamma}(M,1/M)$ and $Y\sim\mr{ln}\mathcal{N}(\hat{\mu},\hat{\sigma}^2)$ are independent Gamma and Lognormal distributed random variables, respectively, then $Z=XY\sim GL(M,\tfrac{1}{M},\hat{\mu},\hat{\sigma}^2)$ is defined as a composite GL distributed random variable with $\mathbb{E}\{Z\}=e^{\hat{\mu}+0.5\hat{\sigma}^2}$. As the exact PDF is intractable for deriving the corresponding CDF and Laplace transform, a good approximation is provided by \cite[Lemma 1]{7857035}.

The corresponding approximated CDF is given as
\begin{align}
     F_Z(z)\approx\tfrac{1}{\Gamma(M)\sum_{t=1}^{T}w_t}\sum_{t=1}^{T}w_t\gamma\left(M,zMe^{-(\sqrt{2}\hat{\sigma}\nu_t+\hat{\mu})}\right),
\end{align}
where $\gamma(s,x)=\int_0^xt^{s-1}e^{-t}\mr{d}t$ is the incomplete gamma function. $T$, $w_t$, and $\nu_t$ are the number of gamma distribution samples, weight, and abscissas factors for the Gaussian-Hermite integration, respectively, whose values can be obtained from \cite{Salzer1952TableOT}. 

The Laplace transform is approximated as
\begin{align}
    \mathscr{L}_Z(s)=\mathbb{E}\left\{e^{-sz}\right\}\approx\tfrac{1}{\sum_{t=1}^{T}w_t}\sum_{t=1}^{T}\tfrac{w_tM^Me^{-M(\sqrt{2}\hat{\sigma}\nu_t+\hat{\mu})}}{\left(s+Me^{-(\sqrt{2}\hat{\sigma}\nu_t+\hat{\mu})}\right)^M}.\label{GLMGF}
\end{align}

\newcounter{myempeqncn}
\begin{figure*}[htbp]
\normalsize
\setcounter{equation}{32}
\begin{align}
    &\mathbb{E}\left\{e^{-\mathcal{G}_{\mr{s},t}r_\mr{s,L}^{\alpha_{\mr{s},\mr{L}}}\hat{I}_\mr{s,a}^{\mr{L}}}\right\}\overset{(c)}{\approx}\mathbb{E}\left\{\prod\limits_{r_\mr{a,\mr{L}}>r_\mr{s,L}} \tfrac{1}{\sum_{p=1}^{T}w_p}\sum_{p=1}^{T}\tfrac{w_pM_{\mr{s},\mr{L}}^{M_{\mr{s},\mr{L}}}e^{-M_{\mr{s},\mr{L}}(\sqrt{2}\hat{\sigma}_{\mr{s},\mr{L}}\nu_p+\hat{\mu})}}{\left(\mathcal{G}_{\mr{s},t}\tfrac{P_\mr{s}N^\mr{s}_\mr{T}\left(N^\mr{u}_\mr{R}\right)^2r_\mr{s,L}^{\alpha_{\mr{s},\mr{L}}}}{KN_\mr{s}r^{\alpha_{\mr{s},\mr{L}}}_\mr{a,L}}p_\mr{BF}(N^\mr{s}_\mr{T},N^\mr{u}_\mr{R},N_\mr{s})+M_{\mr{s},\mr{L}}e^{-(\sqrt{2}\hat{\sigma}_{\mr{s},\mr{L}}\nu_p+\hat{\mu})}\right)^{M_{\mr{s},\mr{L}}}}\right\}\notag\\&\overset{(d)}{=}e^{-2\pi\int_{r_\mr{s,L}}^{R_0}\lambda_{\mr{s},\mr{L}}(r_\mr{a,\mr{L}})r_\mr{a,\mr{L}}\left[1-\tfrac{1}{\sum_{p=1}^{T}w_p}\sum_{p=1}^{T}\tfrac{w_pM_{\mr{s},\mr{L}}^{M_{\mr{s},\mr{L}}}e^{-M_{\mr{s},\mr{L}}(\sqrt{2}\hat{\sigma}_{\mr{s},\mr{L}}\nu_p+\hat{\mu})}}{\left(\mathcal{G}_{\mr{s},t}\tfrac{P_\mr{s}N^\mr{s}_\mr{T}\left(N^\mr{u}_\mr{R}\right)^2r_\mr{s,L}^{\alpha_{\mr{s},\mr{L}}}}{KN_\mr{s}r^{\alpha_{\mr{s},\mr{L}}}_\mr{a,L}}p_\mr{BF}(N^\mr{s}_\mr{T},N^\mr{u}_\mr{R},N_\mr{s})+M_{\mr{s},\mr{L}}e^{-(\sqrt{2}\hat{\sigma}_{\mr{s},\mr{L}}\nu_p+\hat{\mu})}\right)^{M_{\mr{s},\mr{L}}}}\right]\mr{d}r_\mr{a,\mr{L}}}\notag\\
&\overset{(e)}{=}\prod\limits_{i\in \atop p_\mr{BF}(N^\mr{s}_\mr{T},N^\mr{u}_\mr{R},N_\mr{s})} e^{-2\pi\int_{r_\mr{s,L}}^{R_0}\lambda_{\mr{s},\mr{L}}(r_\mr{a,\mr{L}})r_\mr{a,\mr{L}}\left[1-\tfrac{1}{\sum_{p=1}^{T}w_p}\sum_{p=1}^{T}\tfrac{w_pM_{\mr{s},\mr{L}}^{M_{\mr{s},\mr{L}}}e^{-M_{\mr{s},\mr{L}}(\sqrt{2}\hat{\sigma}_{\mr{s},\mr{L}}\nu_p+\hat{\mu})}}{\left(\mathcal{G}_{\mr{s},t}\tfrac{P_\mr{s}N^\mr{s}_\mr{T}\left(N^\mr{u}_\mr{R}\right)^2r_\mr{s,L}^{\alpha_{\mr{s},\mr{L}}}}{KN_\mr{s}r^{\alpha_{\mr{s},\mr{L}}}_\mr{a,L}}b_i+M_{\mr{s},\mr{L}}e^{-(\sqrt{2}\hat{\sigma}_{\mr{s},\mr{L}}\nu_p+\hat{\mu})}\right)^{M_{\mr{s},\mr{L}}}}\right]c_i\mr{d}r_\mr{a,\mr{L}}},\label{33}
\end{align}
\hrulefill
\setcounter{equation}{\value{myempeqncn}}
\end{figure*}

\section{Proof of Proposition 2}\label{app2}
A typical UE is associated with its nearest LoS IAB-node rather than the nearest gNB and NLoS IAB-node if the nearest LoS IAB-node can provide the highest long-term averaged biased-received-desired-signal power, given as
\begin{align}
    &\mathcal{A}_{\mr{s},\mr{L}}=\mathbb{P}\left(\tfrac{P_\mr{s}N_\mr{T}^\mr{s}\left(N_\mr{R}^\mr{u}\right)^2T_\mr{s}}{KN^2_\mr{s}}\mathbb{E}\left\{\widetilde{h}^{\mathbf{0}\mathbf{x}^\mr{L}_\mr{s}}\right\}R_{\mathbf{x}^\mr{L}_\mr{s}\mathbf{0}}^{-\alpha_{\mr{s},\mr{L}}}\geq\tfrac{P_\mr{m}N_\mr{T}^\mr{m}\left(N_\mr{R}^\mr{u}\right)^2T_\mr{m}}{KN_\mr{m}^2}\right.\notag\\&\;\quad\;\;\quad\times\mathbb{E}\left\{\widetilde{h}^{\mathbf{0}\mathbf{x}_\mr{m}}\right\} R_{\mathbf{x}_\mr{m}\mathbf{0}}^{-\alpha_\mr{m}}\bigcap\tfrac{P_\mr{s}N^\mr{s}_\mr{T}\left(N_\mr{R}^\mr{u}\right)^2T_\mr{s}}{KN^2_\mr{s}}\mathbb{E}\left\{\widetilde{h}^{\mathbf{0}\mathbf{x}^\mr{L}_\mr{s}}\right\}R_{\mathbf{x}^\mr{L}_\mr{s}\mathbf{0}}^{-\alpha_{\mr{s},\mr{L}}}\notag\\&\left.\;\quad\;\;\quad\geq\tfrac{P_\mr{s}N_\mr{T}^\mr{s}\left(N_\mr{R}^\mr{u}\right)^2T_\mr{s}}{KN^2_\mr{s}}\mathbb{E}\left\{\widetilde{h}^{\mathbf{0}\mathbf{x}^\mr{N}_\mr{s}}\right\}R_{\mathbf{x}^\mr{N}_\mr{s}\mathbf{0}}^{-\alpha_{\mr{s},\mr{N}}}\right)\notag\\&\overset{(a)}{=}\mathbb{P}\left(\tfrac{P_\mr{s}N^\mr{s}_\mr{T}\left(N_\mr{R}^\mr{u}\right)^2T_\mr{s}}{KN^2_\mr{s}}e^{-0.1\beta\mr{ln}10+\tfrac{(0.1\zeta_{\mr{s},\mr{L}}\mr{ln}10)^2}{2}}R_{\mathbf{x}_\mr{s}^\mr{L}\mathbf{0}}^{-\alpha_{\mr{s},\mr{L}}}\right.\geq R_{\mathbf{x}_\mr{m}\mathbf{0}}^{-\alpha_\mr{m}}\notag\\&\;\quad\times\tfrac{P_\mr{m}N^\mr{m}_\mr{T}\left(N_\mr{R}^\mr{u}\right)^2T_\mr{m}}{KN_\mr{m}^2}e^{-0.1\beta\mr{ln}10+\tfrac{(0.1\zeta_\mr{m}\mr{ln}10)^2}{2}}\bigcap\tfrac{P_\mr{s}N^\mr{s}_\mr{T}\left(N_\mr{R}^\mr{u}\right)^2T_\mr{s}}{KN^2_\mr{s}}\notag\\&\quad\;\times e^{-0.1\beta\mr{ln}10+\tfrac{(0.1\zeta_{\mr{s},\mr{L}}\mr{ln}10)^2}{2}}R_{\mathbf{x}_\mr{s}^\mr{L}\mathbf{0}}^{-\alpha_{\mr{s},\mr{L}}}\geq\tfrac{P_\mr{s}N^\mr{s}_\mr{T}\left(N_\mr{R}^\mr{u}\right)^2T_\mr{s}}{KN^2_\mr{s}}R_{\mathbf{x}_\mr{s}^\mr{N}\mathbf{0}}^{-\alpha_{\mr{s},\mr{N}}}\notag\\&\quad\;\left.\times e^{-0.1\beta\mr{ln}10+\tfrac{(0.1\zeta_{\mr{s},\mr{N}}\mr{ln}10)^2}{2}}\right)\notag\\
    &\overset{(b)}{=}\int_0^\infty\mathbb{P}\left(R_{\mathbf{x}_\mr{m}\mathbf{0}}\geq r_{\mr{s},\mr{L}}^{\tfrac{\alpha_{\mr{s},\mr{L}}}{\alpha_\mr{m}}}\Delta_{\mr{L},1}\right)\mathbb{P}\left(R_{\mathbf{x}_\mr{s}^\mr{N}\mathbf{0}}\geq r_{\mr{s},\mr{L}}^{\tfrac{\alpha_{\mr{s},\mr{L}}}{\alpha_{\mr{s},\mr{N}}}}\Delta_{\mr{L},2}\right)\notag\\&\quad\;\times f_{R_{\mathbf{x}_\mr{s}^\mr{L}\mathbf{0}}}(r_{\mr{s},\mr{L}})\mr{d}r_{\mr{s},\mr{L}},
\end{align}
where $(a)$ is derived by substituting the mean power of the composite fading. $(b)$ is obtained by taking the average of $R_{\mathbf{x}_\mr{s}^\mr{L}\mathbf{0}}=r_{\mr{s},\mr{L}}$. By using Lemma 1 and Lemma 2, we can complete the proof. The proof of $\mathcal{A}_{\mr{s},\mr{N}}$ can be provided in a similar manner.

\section{A Useful Lemma}\label{lem3}
A useful lemma shown below can support the derivation of $\mr{SINR}$ coverage. With the following lemma, one can easily deploy the PGFL of PPP to estimate the mean interference from gNBs in the backhaul link and the gNB-associated access link.

\textit{Lemma 3: Given the density of the MHCPP gNB as $\lambda_\mr{m}$ with hard-core distance $\xi$, it can be approximated by thinning the density of its parent PPP $\widetilde{\lambda}_\mr{m}$ with $\rho_\mr{M}(r_0)$, where $\rho_\mr{M}(r_0)$ is the conditional thinning Palm probability with $r_0>0$ being the distance between the two gNBs, which is expressed as
\setcounter{equation}{30}
\begin{align}
\rho_\mr{M}(r_0)=\left\{
\begin{aligned}
&\tfrac{2}{\widetilde{\lambda}_\mr{m}(\kappa-\pi \xi^2)}\left[1-\tfrac{\pi \xi^2\widetilde{\lambda}_\mr{m}\left(1-e^{-\widetilde{\lambda}_\mr{m}\kappa}\right)}{\widetilde{\lambda}_\mr{m}\kappa\left(1-e^{-\pi \xi^2\widetilde{\lambda}_\mr{m}}\right)}\right] \;\xi\leq r_0<2\xi\\
&\rho \quad\;r_0\geq2\xi\\
&0 \quad\text{otherwise},
\end{aligned}
\right.
\end{align}
with $\kappa=2\pi\xi^2-2\xi^2cos^{-1}\left(\tfrac{r_0}{2\xi}\right)+r_0\sqrt{\xi^2-\tfrac{r_0^2}{4}}$.}
\begin{IEEEproof}
A proof can be found in \cite{9350211}.
\end{IEEEproof}

\section{Solution of Theorem 1}\label{appB}
\subsection{Solution of $\mathbb{P}\left(\mr{SINR}_{\mathbf{0}\mathbf{x}_\mr{s}^{\mr{L}}}>\tau\right)$}
The $\mr{SINR}$ coverage for the LoS IAB-node associated access link is expressed as
\setcounter{equation}{31}
\begin{align}
&\mathbb{P}\left(\mr{SINR}_{\mathbf{0}\mathbf{x}_\mr{s}^{\mr{L}}}>\tau\right)=\left(1-\tfrac{1}{N^\mr{s}_\mr{T}}\right)^{N_\mr{s}-1}\int_0^{R_0}\mathbb{P}\left(\widetilde{h}^{\mathbf{0}\mathbf{x}_\mr{s}^{\mr{L}}}>\mathcal{G}_\mr{s}\right.\notag\\&\left.\quad\;\times\underbrace{r_\mr{s,L}^{\alpha_{\mr{s},\mr{L}}}\left(\hat{I}_\mr{m,a}+\hat{I}_\mr{s,a}^{\mr{L}}+\hat{I}_\mr{s,a}^{\mr{N}}+\sigma_\mr{n}^2N^\mr{u}_\mr{R}\right)}_{I}\right)\hat{f}_{R_{\mathbf{x}_\mr{s}^\mr{L}\mathbf{0}}}(r_\mr{s,L})\mr{d}r_\mr{s,L}\notag\\&\overset{(a)}{\approx} \left(1-\tfrac{1}{N^\mr{s}_\mr{T}}\right)^{N_\mr{s}-1}\int_0^{R_0}\left[1-\mathbb{E}\left\{\tfrac{1}{\Gamma(M_{\mr{s},\mr{L}})\sum_{t=1}^{T}w_t}\sum_{t=1}^{T}w_t\right.\right.\notag\\&\quad\;\times\left.\left.\gamma\left(M_{\mr{s},\mr{L}},{\mathcal{G}_{\mr{s},t}I}\right)\right\}\right]\hat{f}_{R_{\mathbf{x}_\mr{s}^\mr{L}\mathbf{0}}}(r_\mr{s,L})\mr{d}r_\mr{s,L}\notag\\&\overset{(b)}{=}\left(1-\tfrac{1}{N^\mr{s}_\mr{T}}\right)^{N_\mr{s}-1}\sum_{t=1}^{T}\sum_{n=0}^{M_{\mr{s},\mr{L}}-1}\frac{w_t}{n!\sum_{t=1}^{T}w_t}\notag\\&\quad\;\times\int_0^{R_0}\mathbb{E}\left\{\left(\mathcal{G}_{\mr{s},t}I\right)^ne^{-\mathcal{G}_{\mr{s},t}I}\right\}\hat{f}_{R_{\mathbf{x}_\mr{s}^\mr{L}\mathbf{0}}}(r_\mr{s,L})\mr{d}r_\mr{s,L}\notag\\&=\left(1-\tfrac{1}{N^\mr{s}_\mr{T}}\right)^{N_\mr{s}-1}\sum_{t=1}^{T}\sum_{n=0}^{M_{\mr{s},\mr{L}}-1}\frac{w_t(-\mathcal{G}_{\mr{s},t})^n}{n!\sum_{t=1}^{T}w_t}\frac{\mr{d}^n}{\mr{d}\mathcal{G}_{\mr{s},t}^n}\notag\\&\quad\;\times\left(\int_0^{R_0} \mathbb{E}\left\{e^{-\mathcal{G}_{\mr{s},t}r_\mr{s,L}^{\alpha_{\mr{s},\mr{L}}}\hat{I}_\mr{s,a}^{\mr{L}}}\right\}\mathbb{E}\left\{e^{-\mathcal{G}_{\mr{s},t}r_\mr{s,L}^{\alpha_{\mr{s},\mr{L}}}\hat{I}_\mr{s,a}^{\mr{N}}}\right\}\right.\notag\\&\quad\;\left.\times\mathbb{E}\left\{e^{-\mathcal{G}_{\mr{s},t}r_\mr{s,L}^{\alpha_{\mr{s},\mr{L}}}\hat{I}_\mr{m,a}}\right\}e^{-\mathcal{G}_{\mr{s},t}r_\mr{s,L}^{\alpha_{\mr{s},\mr{L}}}\sigma_\mr{n}^2N^\mr{u}_\mr{R}}\hat{f}_{R_{\mathbf{x}_\mr{s}^\mr{L}\mathbf{0}}}(r_\mr{s,L})\mr{d}r_\mr{s,L}\right)\label{30},
\end{align}
where $\mathcal{G}_\mr{s}=\tfrac{\tau KN_\mr{s}^2\left(1+\tfrac{1}{1.5\cdot2^{2q}}\right)}{P_\mr{s}N^\mr{s}_\mr{T}\left(N^\mr{u}_\mr{R}\right)^2\left(1-\tfrac{\tau}{1.5\cdot2^{2q}}\right)}$. $\mathcal{G}_{\mr{s},t}=\mathcal{G}_\mr{s} M_{\mr{s},\mr{L}}e^{-(\sqrt{2}\hat{\sigma}_{\mr{s},\mr{L}}\nu_t+\hat{\mu})}$ with $\hat{\mu}=-0.1\beta\mr{ln}10$ and $\hat{\sigma}_{\mr{s},\mr{L}}=0.1\zeta_{\mr{s},\mr{L}}\mr{ln}10$. $(a)$ is derived from the approximated CDF of the composite GL distribution and $(b)$ can be referred to \cite[Eq.(8.4.8)]{hand}.

Next, we present the mean interference from LoS IAB-nodes in \eqref{33} shown at the top of the page, where $(c)$ is derived according to the approximated Laplace transform of the composite GL distribution. After applying the PGFL of PPP, we obtain $(d)$. In $(e)$, we explore the independence of different beamforming gain scenarios, $b_i$ takes one of the beamforming gains from \eqref{BFpenlty}, and $c_i$ takes the corresponding probability. 

Then, the mean interference from NLoS IAB-nodes $\mathbb{E}\left\{e^{-\mathcal{G}_{\mr{s},t}r_\mr{s,L}^{\alpha_{\mr{s},\mr{L}}}\hat{I}_\mr{s,a}^{\mr{N}}}\right\}$ can be obtained by a similar step in \eqref{33} with integration  from ${r_\mr{s,L}^{\tfrac{\alpha_{\mr{s},\mr{L}}}{\alpha_{\mr{s},\mr{N}}}}\Delta_{\mr{L},2}}$ to ${R_0}$.

Finally, we derive the mean interference from the gNBs. Because of the lack of knowledge of the PGFL of MHCPP, we assume gNBs as a virtual PPP with a density of ${\lambda}_\mr{m}$ to approximate the mean interference, which has been verified to provide a good approximation in \cite{5934671}. Consequently, with the help of the PGFL of PPP, we have \eqref{34} shown at the top of the next page,
\newcounter{myempeqnc}
\begin{figure*}[htbp]
\normalsize
\setcounter{equation}{33}
\begin{align}
   &\mathbb{E}\left\{e^{-\mathcal{G}_{\mr{s},t}r_\mr{s,L}^{\alpha_{\mr{s},\mr{L}}}\hat{I}_\mr{m,a}}\right\}\approx\prod\limits_{i\in \atop p_\mr{BF}(N^\mr{m}_\mr{T},N^\mr{u}_\mr{R},N_\mr{m})} e^{-2\pi\int_{r_\mr{s,L}^{\tfrac{\alpha_{\mr{s},\mr{L}}}{\alpha_\mr{m}}}\Delta_{\mr{L},1}}^{R_0}{\lambda}_\mr{m}\left[1-\tfrac{1}{\sum_{p=1}^{T}w_p}\sum_{p=1}^{T}\tfrac{w_pM_\mr{m}^{M_\mr{m}}e^{-M_\mr{m}(\sqrt{2}\hat{\sigma}_\mr{m}\nu_p+\hat{\mu})}}{\left(\mathcal{G}_{\mr{s},t}\tfrac{P_\mr{m}N_\mr{T}^\mr{m}\left(N^\mr{u}_\mr{R}\right)^2r_\mr{s,L}^{\alpha_{\mr{s},\mr{L}}}}{KN_\mr{m}r^{\alpha_\mr{m}}_\mr{a}}b_i+M_\mr{m}e^{-(\sqrt{2}\hat{\sigma}_\mr{m}\nu_p+\hat{\mu})}\right)^{M_\mr{m}}}\right]c_ir_\mr{a}\mr{d}r_\mr{a}},\label{34}
\end{align}
\setcounter{equation}{\value{myempeqnc}}
\end{figure*}
\newcounter{myempeqn}
\begin{figure*}[htbp]
\normalsize
\setcounter{equation}{34}
\begin{align}
  &\mathbb{E}\left\{e^{-\mathcal{G}_{\mr{b},t}r^{\alpha_\mr{m}}{I}_\mr{m,b}}\right\}\approx\prod\limits_{i\in \atop p_\mr{BF}(N^\mr{m}_\mr{T},N^\mr{s}_\mr{R},N_\mr{m})} e^{-\int_0^{2\pi}\int_{\xi}^{R_0}\widetilde{\lambda}_\mr{m}\rho_\mr{M}(r_0)\left[1-\tfrac{1}{\sum_{p=1}^{T}w_p}\sum_{p=1}^{T}\tfrac{w_pM_\mr{m}^{M_\mr{m}}e^{-M_\mr{m}(\sqrt{2}\hat{\sigma}_\mr{m}\nu_p+\hat{\mu})}}{\left(\mathcal{G}_{\mr{b},t}\tfrac{P_\mr{m}N^\mr{m}_\mr{T}\left(N^\mr{s}_\mr{R}\right)^2r^{\alpha_\mr{m}}}{KN_\mr{m}r_\mr{b}^{\alpha_\mr{m}}}b_i+M_\mr{m}e^{-(\sqrt{2}\hat{\sigma}_\mr{m}\nu_p+\hat{\mu})}\right)^{M_\mr{m}}}\right]c_ir_0\mr{d}r_0\mr{d}\vartheta},\label{39}
\end{align}
\hrulefill
\setcounter{equation}
{\value{myempeqn}}
\end{figure*}
where $\hat{\sigma}_\mr{m}=0.1\zeta_\mr{m}\mr{ln}10$. The solution for $\mathbb{P}\left(\mr{SINR}_{\mathbf{0}\mathbf{x}_\mr{s}^{\mr{N}}}>\tau\right)$ can be similarly derived.

\subsection{Solution of $\mathbb{P}\left(\mr{SINR}_{\hat{\mathbf{0}}\mathbf{x}_\mr{m}}>\tau\right)$}\label{111}
Likewise, the $\mr{SINR}$ coverage for the typical backhaul link $\mathbb{P}\left(\mr{SINR}_{\hat{\mathbf{0}}\mathbf{x}_\mr{m}}>\tau\right)$ can be approximated in a manner similar to \eqref{30}. The mean interference from LoS/NLoS IAB-nodes can be derived from a similar step in \eqref{33} with an integration limit from 0 to ${R_0}$ because the typical IAB-node can only be served by the gNB.

As for the mean interference from other gNBs, we still use the PGFL of PPP to make the approximation, which is given in \eqref{39} shown at the top of the next page, where $\mathcal{G}_\mr{b}=\tfrac{\tau KN_\mr{m}^2\left(1+\tfrac{1}{1.5\cdot2^{2q}}\right)}{P_\mr{m}N^\mr{m}_\mr{T}\left(N^\mr{s}_\mr{R}\right)^2\left(1-\tfrac{\tau}{1.5\cdot2^{2q}}\right)}$, $\mathcal{G}_{\mr{b},t}=\mathcal{G}_\mr{b} M_\mr{m}e^{-(\sqrt{2}\hat{\sigma}_\mr{m}\nu_t+\hat{\mu})}$, $r_\mr{b}=\sqrt{r^2+r_0^2-2rr_0\cos(\vartheta)}$, $\rho_\mr{M}(r_0)$ is the thinning probability derived in \cite[Lemma 1]{9350211}.

\subsection{Solution of $\mathbb{P}\left(\mr{SINR}_{\mathbf{0}\mathbf{x}_\mr{m}}>\tau\right)$}
Similarly, the $\mr{SINR}$ coverage for the gNB-associated access $\mathbb{P}\left(\mr{SINR}_{{\mathbf{0}}\mathbf{x}_\mr{m}}>\tau\right)$ can be approximated by following the steps in \eqref{30}. The mean interference from LoS/NLoS IAB-nodes follows the similar way as in \eqref{33}, with the integration limit from ${r_\mr{m}^{\tfrac{\alpha_\mr{m}}{\alpha_{\mr{s},j}}}\Delta_{j}}$ to ${R_0}$, where $j\in\{\mr{L},\mr{N}\}$. The mean interference from other gNBs can be approximated in a manner similar to that in \eqref{39}.

\section{Solution of Theorem 3}
\label{D}
According to \cite[Lemma 2]{9258892}, we have
\setcounter{equation}{35}
\begin{align}
   &\bar{\mathcal{R}}_\mr{m}=\mathbb{E}\{W\log_2(1+\mr{SINR}_{\mathbf{0}\mathbf{x}_\mr{m}})\}=\left(1-\tfrac{1}{N^\mr{m}_\mr{T}}\right)^{N_\mr{m}-1}\notag\\&\times\int_0^{R_0}\frac{W}{\ln{2}}\int_0^\infty\mathbb{E}\left\{e^{-z\left(\tfrac{{I}_\mr{m,a}+{I}_\mr{s,a}^{\mr{L}}+{I}_\mr{s,a}^{\mr{N}}}{\sigma_\mr{n}^2N^\mr{u}_\mr{R}}+\tfrac{\tfrac{1}{1.5\cdot2^{2q}}G_\mr{a,m}}{\left(1+\tfrac{1}{1.5\cdot2^{2q}}\right)\sigma_\mr{n}^2N^\mr{u}_\mr{R}}\right)}\right\}\notag\\&\times\left(1-\mathbb{E}\left\{e^{-z\tfrac{G_\mr{a,m}}{\left(1+\tfrac{1}{1.5\cdot2^{2q}}\right)\sigma_\mr{n}^2N^\mr{u}_\mr{R}}}\right\}\right)\frac{e^{-z}}{z}\mr{d}z\hat{{f}}_{R_{\mathbf{x}_\mr{m}{\mathbf{0}}}}(r)\mr{d}r.
\end{align}
Similar to the process in Appendix-\ref{appB}, we can obtain the solution by leveraging the approximated Laplace transform of the composite GL distribution in \eqref{GLMGF} and the PGFL of PPP. Similarly, $\bar{\mathcal{R}}_{\mr{s},\mr{L}}$, $\bar{\mathcal{R}}_{\mr{s},\mr{N}}$ and $\bar{\mathcal{R}}_\mr{b}$ can be derived easily.

\bibliographystyle{IEEEtran}
\bibliography{IEEEabrv.bib,Ref.bib}

\begin{thebibliography}{10}
\providecommand{\url}[1]{#1}
\csname url@samestyle\endcsname
\providecommand{\newblock}{\relax}
\providecommand{\bibinfo}[2]{#2}
\providecommand{\BIBentrySTDinterwordspacing}{\spaceskip=0pt\relax}
\providecommand{\BIBentryALTinterwordstretchfactor}{4}
\providecommand{\BIBentryALTinterwordspacing}{\spaceskip=\fontdimen2\font plus
\BIBentryALTinterwordstretchfactor\fontdimen3\font minus
  \fontdimen4\font\relax}
\providecommand{\BIBforeignlanguage}[2]{{%
\expandafter\ifx\csname l@#1\endcsname\relax
\typeout{** WARNING: IEEEtran.bst: No hyphenation pattern has been}%
\typeout{** loaded for the language `#1'. Using the pattern for}%
\typeout{** the default language instead.}%
\else
\language=\csname l@#1\endcsname
\fi
#2}}
\providecommand{\BIBdecl}{\relax}
\BIBdecl

\bibitem{mythesis}
\BIBentryALTinterwordspacing
J.~{Zhang}, ``In-band-full-duplex integrated access and backhaul enabled next
  generation wireless networks,'' Ph.D. dissertation, Inst. Imaging, Data,
  Commun., Univ. Edinburgh, Edinburgh, UK, Oct. 2023. [Online]. Available:
  \url{https://era.ed.ac.uk/handle/1842/41226}
\BIBentrySTDinterwordspacing

\bibitem{MAG}
J.~{Zhang}, N.~{Garg}, M.~{Holm}, and T.~{Ratnarajah}, ``Design of full duplex
  millimeter-wave integrated access and backhaul networks,'' \emph{{IEEE}
  Wireless Commun. Mag.}, vol.~28, no.~1, pp. 60--67, Feb. 2021.

\bibitem{tong}
T.~{Zhang}, S.~{Biswas}, and T.~{Ratnarajah}, ``An analysis on wireless edge
  caching in in-band full-duplex {FR2-IAB} networks,'' \emph{IEEE Access},
  vol.~8, pp. 164\,987--165\,002, Sep. 2020.

\bibitem{7448873}
S.~{Park}, A.~{Alkhateeb}, and R.~W. {Heath}, ``Dynamic subarrays for hybrid
  precoding in wideband mm{W}ave {MIMO} systems,'' \emph{{IEEE} Trans. Wireless
  Commun.}, vol.~16, no.~5, pp. 2907--2920, May 2017.

\bibitem{7434599}
M.~N. {Kulkarni}, A.~{Ghosh}, and J.~G. {Andrews}, ``A comparison of {MIMO}
  techniques in downlink millimeter wave cellular networks with hybrid
  beamforming,'' \emph{{IEEE} Trans. Commun.}, vol.~64, no.~5, pp. 1952--1967,
  March 2016.

\bibitem{8246856}
Z.~{Xiao}, P.~{Xia}, and X.~{Xia}, ``Full-duplex millimeter-wave
  communication,'' \emph{{IEEE} Wireless Commun. Mag.}, vol.~24, no.~6, pp.
  136--143, Dec. 2017.

\bibitem{7562572}
A.~C. Cirik, S.~Biswas, S.~Vuppala, and T.~Ratnarajah, ``Beamforming design for
  full-duplex {MIMO} interference channels–{QoS} and energy-efficiency
  considerations,'' \emph{{IEEE} Trans. Commun.}, vol.~64, no.~11, pp.
  4635--4651, Nov. 2016.

\bibitem{luo}
H.~Luo, A.~Bishnu, and T.~Ratnarajah, ``Design and analysis of in-band
  full-duplex private {5G} networks using {FR2} band,'' \emph{IEEE Access},
  vol.~9, pp. 166\,886--166\,905, Dec. 2021.

\bibitem{9431171}
A.~Bishnu, M.~Holm, and T.~Ratnarajah, ``Performance evaluation of full-duplex
  {IAB} multi-cell and multi-user network for {FR2} band,'' \emph{IEEE Access},
  vol.~9, pp. 72\,269--72\,283, May 2021.

\bibitem{9500615}
J.~Zhang, H.~Luo, N.~Garg, M.~Holm, and T.~Ratnarajah, ``Design and analysis of
  mm{W}ave full-duplex integrated access and backhaul networks,'' in
  \emph{Proc. IEEE ICC}, June 2021.

\bibitem{Zhan2012}
J.~{Zhang}, H.~{Luo}, N.~{Garg}, A.~{Bishnu}, M.~{Holm}, and T.~{Ratnarajah},
  ``Design and analysis of wideband in-band-full-duplex {FR2-IAB} networks,''
  \emph{{IEEE} Trans. Wireless Commun.}, vol.~21, no.~6, pp. 4183--4196, June
  2022.

\bibitem{3gpp}
3GPP, ``{NR}; {S}tudy on {I}ntegrated {A}ccess and {B}ackhaul,'' \emph{TR
  38.874 (Rel. 16)}, Dec. 2018.

\bibitem{8252876}
J.~García-Rois, R.~Banirazi, F.~J. González-Castaño, B.~Lorenzo, and J.~C.
  Burguillo, ``Delay-aware optimization framework for proportional flow delay
  differentiation in millimeter-wave backhaul cellular networks,'' \emph{{IEEE}
  Trans. Commun.}, vol.~66, no.~5, pp. 2037--2051, May 2018.

\bibitem{8493520}
C.~{Saha}, M.~{Afshang}, and H.~S. {Dhillon}, ``Bandwidth partitioning and
  downlink analysis in millimeter wave integrated access and backhaul for
  {5G},'' \emph{{IEEE} Trans. Wireless Commun.}, vol.~17, no.~12, pp.
  8195--8210, Oct. 2018.

\bibitem{9040265}
M.~{Polese}, M.~{Giordani}, T.~{Zugno}, A.~{Roy}, S.~{Goyal}, D.~{Castor}, and
  M.~{Zorzi}, ``Integrated access and backhaul in {5G} mmwave networks:
  Potential and challenges,'' \emph{{IEEE} Commun. Mag.}, vol.~58, no.~3, pp.
  62--68, March 2020.

\bibitem{8882288}
C.~{Saha} and H.~S. {Dhillon}, ``Millimeter wave integrated access and backhaul
  in {5G}: Performance analysis and design insights,'' \emph{{IEEE} J. Select.
  Areas Commun.}, vol.~37, no.~12, pp. 2669--2684, Dec. 2019.

\bibitem{7110547}
S.~{Singh}, M.~N. {Kulkarni}, A.~{Ghosh}, and J.~G. {Andrews}, ``Tractable
  model for rate in self-backhauled millimeter wave cellular networks,''
  \emph{{IEEE} J. Select. Areas Commun.}, vol.~33, no.~10, pp. 2196--2211, May
  2015.

\bibitem{6515173}
T.~S. {Rappaport}, S.~{Sun}, R.~{Mayzus}, H.~{Zhao}, Y.~{Azar}, K.~{Wang},
  G.~N. {Wong}, J.~K. {Schulz}, M.~{Samimi}, and F.~{Gutierrez}, ``Millimeter
  wave mobile communications for 5{G} cellular: It will work!'' \emph{IEEE
  Access}, vol.~1, pp. 335--349, May 2013.

\bibitem{6932503}
T.~{Bai} and R.~W. {Heath}, ``Coverage and rate analysis for millimeter-wave
  cellular networks,'' \emph{{IEEE} Trans. Wireless Commun.}, vol.~14, no.~2,
  pp. 1100--1114, Oct. 2015.

\bibitem{6287527}
H.-S. Jo, Y.~J. Sang, P.~Xia, and J.~G. Andrews, ``Heterogeneous cellular
  networks with flexible cell association: A comprehensive downlink {SINR}
  analysis,'' \emph{{IEEE} Trans. Wireless Commun.}, vol.~11, no.~10, pp.
  3484--3495, Oct. 2012.

\bibitem{8335329}
Y.~{Zhu}, G.~{Zheng}, and M.~{Fitch}, ``Secrecy rate analysis of {UAV}-enabled
  mm{W}ave networks using {M}atérn hardcore point processes,'' \emph{{IEEE} J.
  Select. Areas Commun.}, vol.~36, no.~7, pp. 1397--1409, April 2018.

\bibitem{8648502}
S.~S. Kalamkar and M.~Haenggi, ``Simple approximations of the {SIR} meta
  distribution in general cellular networks,'' \emph{{IEEE} Trans. Commun.},
  vol.~67, no.~6, pp. 4393--4406, Feb. 2019.

\bibitem{9350211}
J.~Lyu and H.-M. Wang, ``Secure {UAV} random networks with minimum safety
  distance,'' \emph{{IEEE} Trans. Veh. Technol.}, vol.~70, no.~3, pp.
  2856--2861, March 2021.

\bibitem{9258892}
H.~{He}, S.~{Biswas}, P.~{Aquilina}, T.~{Ratnarajah}, and J.~{Yang},
  ``Performance analysis of multi-cell full-duplex cellular networks,''
  \emph{IEEE Access}, vol.~8, pp. 206\,914--206\,930, Nov. 2020.

\bibitem{7415954}
H.~He, J.~Xue, T.~Ratnarajah, F.~A. Khan, and C.~B. Papadias, ``Modeling and
  analysis of cloud radio access networks using {Matérn} hard-core point
  processes,'' \emph{{IEEE} Trans. Wireless Commun.}, vol.~15, no.~6, pp.
  4074--4087, June 2016.

\bibitem{7817893}
A.~{Sharma}, R.~K. {Ganti}, and J.~K. {Milleth}, ``Joint backhaul-access
  analysis of full duplex self-backhauling heterogeneous networks,''
  \emph{{IEEE} Trans. Wireless Commun.}, vol.~16, no.~3, pp. 1727--1740, Jan.
  2017.

\bibitem{7448962}
A.~Thornburg, T.~Bai, and R.~W. Heath, ``Performance analysis of outdoor
  mm{W}ave ad hoc networks,'' \emph{{IEEE} Trans. Signal Processing}, vol.~64,
  no.~15, pp. 4065--4079, Aug. 2016.

\bibitem{7434598}
M.~N. {Kulkarni}, A.~{Alkhateeb}, and J.~G. {Andrews}, ``A tractable model for
  per user rate in multiuser millimeter wave cellular networks,'' in
  \emph{Proc. 49th Asilomar Conference on Signals, Systems and Computers}, Nov.
  2015.

\bibitem{7857035}
K.~Cho, J.~Lee, and C.~G. Kang, ``Stochastic geometry-based coverage and rate
  analysis under {N}akagami {L}og-normal composite fading channel for downlink
  cellular networks,'' \emph{IEEE Commun. Lett.}, vol.~21, no.~6, pp.
  1437--1440, June 2017.

\bibitem{haenggi_2012}
M.~Haenggi, \emph{Stochastic Geometry for Wireless Networks}.\hskip 1em plus
  0.5em minus 0.4em\relax Cambridge University Press, 2012.

\bibitem{5934671}
------, ``Mean interference in hard-core wireless networks,'' \emph{IEEE
  Commun. Lett.}, vol.~15, no.~8, pp. 792--794, Aug. 2011.

\bibitem{6840343}
T.~{Bai}, R.~{Vaze}, and R.~W. {Heath}, ``Analysis of blockage effects on urban
  cellular networks,'' \emph{{IEEE} Trans. Wireless Commun.}, vol.~13, no.~9,
  pp. 5070--5083, Sep. 2014.

\bibitem{NAKAGAMI19603}
M.~Nakagami, ``The m-distribution—a general formula of intensity distribution
  of rapid fading,'' in \emph{Statistical Methods in Radio Wave
  Propagation}.\hskip 1em plus 0.5em minus 0.4em\relax W. G. Hoffman, Ed,
  Oxford, U.K.: Pergamon, 1960.

\bibitem{6015565}
C.~Zhong, M.~Matthaiou, G.~K. Karagiannidis, and T.~Ratnarajah, ``Generic
  ergodic capacity bounds for fixed-gain {AF} dual-hop relaying systems,''
  \emph{{IEEE} Trans. Veh. Technol.}, vol.~60, no.~8, pp. 3814--3824, Oct.
  2011.

\bibitem{1033686}
A.~M. {Sayeed}, ``Deconstructing multiantenna fading channels,'' \emph{{IEEE}
  Trans. Signal Processing}, vol.~50, no.~10, pp. 2563--2579, Oct. 2002.

\bibitem{6292865}
O.~E. Ayach, R.~W. Heath, S.~Abu-Surra, S.~Rajagopal, and Z.~Pi, ``The capacity
  optimality of beam steering in large millimeter wave {MIMO} systems,'' in
  \emph{Proc. IEEE SPAWC}, June 2012.

\bibitem{antenna}
C.~Balanis, \emph{Antenna Theory}.\hskip 1em plus 0.5em minus 0.4em\relax
  Wiley, 1997.

\bibitem{roberts2020hybrid}
I.~P. Roberts, J.~G. Andrews, and S.~Vishwanath, ``Hybrid beamforming for
  millimeter wave full-duplex under limited receive dynamic range,''
  \emph{{IEEE} Trans. Wireless Commun.}, vol.~20, no.~12, pp. 7758--7772, June
  2021.

\bibitem{5285161}
R.~Giuliano and F.~Mazzenga, ``Exponential effective {SINR} approximations for
  {OFDM/OFDMA}-based cellular system planning,'' \emph{{IEEE} Trans. Wireless
  Commun.}, vol.~8, no.~9, pp. 4434--4439, Oct. 2009.

\bibitem{6213041}
A.~Oborina, M.~Moisio, and V.~Koivunen, ``Performance of mobile {MIMO} {OFDM}
  systems with application to {UTRAN} {LTE} downlink,'' \emph{{IEEE} Trans.
  Wireless Commun.}, vol.~11, no.~8, pp. 2696--2706, Aug. 2012.

\bibitem{6497002}
S.~Singh, H.~S. Dhillon, and J.~G. Andrews, ``Offloading in heterogeneous
  networks: Modeling, analysis, and design insights,'' \emph{{IEEE} Trans.
  Wireless Commun.}, vol.~12, no.~5, pp. 2484--2497, May 2013.

\bibitem{7511676}
A.~Al-Hourani, R.~J. Evans, and S.~Kandeepan, ``Nearest neighbor distance
  distribution in hard-core point processes,'' \emph{IEEE Commun. Lett.},
  vol.~20, no.~9, pp. 1872--1875, Sep. 2016.

\bibitem{8060551}
C.~{Chen}, R.~C. {Elliott}, and W.~A. {Krzymień}, ``Empirical distribution of
  nearest-transmitter distance in wireless networks modeled by {Matérn} hard
  core point processes,'' \emph{{IEEE} Trans. Veh. Technol.}, vol.~67, no.~2,
  pp. 1740--1749, Oct. 2018.

\bibitem{ieee}
\BIBentryALTinterwordspacing
IEEE, ``{IEEE} 802.11ad standard draft {D0.1},'' 2012. [Online]. Available:
  \url{https://www.ieee802.org/11/Reports/tgadupdate.htm}
\BIBentrySTDinterwordspacing

\bibitem{6894456}
A.~I. Sulyman, A.~T. Nassar, M.~K. Samimi, G.~R. Maccartney, T.~S. Rappaport,
  and A.~Alsanie, ``Radio propagation path loss models for {5G} cellular
  networks in the 28 {GH}z and 38 {GH}z millimeter-wave bands,'' \emph{{IEEE}
  Commun. Mag.}, vol.~52, no.~9, pp. 78--86, Sep. 2014.

\bibitem{8337813}
J.~Zhang, L.~Dai, X.~Li, Y.~Liu, and L.~Hanzo, ``On low-resolution {ADCs} in
  practical {5G} millimeter-wave massive {MIMO} systems,'' \emph{{IEEE} Commun.
  Mag.}, vol.~56, no.~7, pp. 205--211, April 2018.

\bibitem{glover2010digital}
I.~Glover and P.~M. Grant, \emph{Digital communications}.\hskip 1em plus 0.5em
  minus 0.4em\relax Pearson Education, 2010.

\bibitem{4534833}
S.~Ahmed, T.~Ratnarajah, M.~Sellathurai, and C.~F.~N. Cowan, ``Iterative
  receivers for {MIMO-OFDM} and their convergence behavior,'' \emph{{IEEE}
  Trans. Veh. Technol.}, vol.~58, no.~1, pp. 461--468, June 2009.

\bibitem{Salzer1952TableOT}
H.~E. Salzer, R.~Zucker, and R.~Capuano, ``Table of the zeros and weight
  factors of the first 20 hermite polynomials,'' \emph{Journal of research of
  the National Bureau of Standards}, vol.~48, pp. 111--116, Feb. 1952.

\bibitem{hand}
F.~Olver, D.~Lozier, R.~Boisvert, and C.~Clark, \emph{NIST Handbook of
  Mathematical Functions}.\hskip 1em plus 0.5em minus 0.4em\relax Cambridge
  University Press, 2010.

\end{thebibliography}

%If you do not have or do not want to include a photo, you can use IEEEbiographynophoto as shown below:

\end{document}